\documentclass{aa}

\renewcommand{\rm}{\mathrm}

\def\({\left(}
\def\r){\right)}

\newcommand*{\nt}{\textrm}
\newcommand*{\dif}{\textrm{d}}

\usepackage{bm}
\usepackage{verbatim}
\usepackage{subcaption}
\usepackage{graphicx}
\usepackage{txfonts}
\usepackage{subcaption}
\usepackage{acronym}
\usepackage[margin=10pt,font=footnotesize,labelfont=bf]{caption}
\usepackage[colorlinks = true,
            linkcolor = blue,
            urlcolor  = blue,
            citecolor = blue,
            anchorcolor = blue]{hyperref}

\usepackage{todonotes}

\begin{document}

   \title{An adapted filter function for density split statistics in weak lensing}

%   \subtitle{}

   \author{Pierre Burger\inst{1}
          \and Peter Schneider\inst{1} 
          \and Vasiliy Demchenko\inst{2}
          \and Joachim Harnois-Deraps\inst{2,3}
          \and \\Catherine Heymans\inst{2,4}
          \and Hendrik Hildebrandt\inst{4}
          \and Sandra Unruh\inst{1}
          }
   \institute{Argelander-Institut f\"ur Astronomie, Auf dem H\"ugel 71, 53121 Bonn, Germany \\ \email{pburger@astro.uni-bonn.de}
\and
Institute for Astronomy, University of Edinburgh, Royal Observatory, Blackford Hill, Edinburgh EH9 3HJ, UK
\and  
Astrophysics Research Institute, Liverpool John Moores University, 146 Brownlow Hill, Liverpool L3 5RF
\and
Astronomisches Institut, Ruhr-Universität Bochum, German Centre for Cosmological Lensing,  Universitätsstr. 150, 44801, Bochum, Germany}

   \date{Received 18 June 2020 / Accepted 25 August 2020}

  \abstract
   {The density split statistics in weak gravitational lensing
     analyses probes the correlation between  regions of different
     (foreground) galaxy number densities and their weak lensing
     signal, which is measured by the shape distortion of background galaxies.}
   {In this paper, we reconsider density split statistics, by
     constructing a new angular filter function that is adapted to the expected relation between the galaxy number density and shear
     pattern, in a way that the filter weighting the galaxy number
     density is matched to the filter that is used to quantify the
     shear signal.}
   {We used the results of numerical ray-tracing simulations,
     specifically through the Millennium Simulation supplemented by a galaxy distribution based on a semi-analytic model, to construct a matched pair of adapted filter functions for the galaxy density and the tangential shear signal. We compared the performance of our new filter to the previously used top-hat filter, applying both to a different and independent set of numerical simulations (SLICS, cosmo-SLICS).}
   {We show that the adapted filter yields a better 
     correlation between the total matter and the galaxy distribution.
     Furthermore, the adapted filter provides a larger signal-to-noise ratio to
     constrain the bias between the total matter and the galaxy
     distribution, and we show that it is, in general, a more sensitive discriminator between different cosmologies, with the exception of cosmologies with very large $\sigma_8$ values. All analyses lead to the conclusion that our adapted filter should be favoured in future density split statistic works.}
   {}

   \keywords{gravitational lensing: weak --methods: statistical -- surveys -- Galaxy: abundances -- (cosmology:) large-sclae structure of Universe
               }

   \maketitle

%
%-------------------------------------------------------------------
\section{Introduction}
 
The large-scale structure (LSS) of the Universe is thought to
originate from initially Gaussian density perturbations, a view
supported by the apparent absence of non-Gaussian features in the
cosmic microwave background \citep[see][]{Aghanim:2018}.
Correspondingly, at early times, these Gaussian perturbations result in a total symmetry in the abundance and amplitude of over- and under-dense regions. As structures evolve, this symmetry breaks so over-densities can grow to very large amplitudes. However, the
fractional density contrast of under-densities is bounded from below.

Studying the matter distribution of the present LSS reveals a wealth of information about the evolution of the Universe. In particular, its distorting effect on the propagation of light from distant galaxies, dubbed cosmic shear,  can be captured by analysing weak lensing surveys. By comparing the results of cosmological models with the observed signal, one can constrain cosmological parameters \citep[see][]{Hildebrandt:2016,DES:2020,Hamana:2020}.

The preferred methods to infer statistical properties of the matter
and galaxy distribution are second- and higher-order
statistics. Two-point correlation functions, or power spectra, measure the variance of density fluctuations as a function of scale. More generally, an $n$-point correlation function describes how probable it is to find a constellation of $n$ connected objects. The advantage of analysing three-point statistics, which are more computationally time-consuming than second-order statistics, is its connection to the skewness of the density distribution resulting from the asymmetric behaviour of over- and under-dense regions. Another advantage of third-order statistics is that they scale differently with cosmological parameters. Hence, by simultaneously investigating second- and higher-order statistics, the power to constrain cosmological parameters increases \citep{Pires:2012,Fu:2014}.

First in \citet{Gruen:2015}, and later in \citet{Gruen:Friedrich:2018} and \citet{Friedrich:Gruen:2018}, a new weak lensing approach to analyse the LSS was introduced, the density split statistics (hereafter DSS), which differs from the usual $n$-point correlation analyses. The idea is to divide the sky into sub-areas of an equal size, according to the foreground (or lens) galaxy density (counts-in-cells, or CiC), and to measure the mean tangential shear, $\gamma_\nt{t}$, around all points within a given sub-area. These sub-areas are defined by quantiles of the galaxy number density field. One expects that around points with a high density of (foreground) galaxies, that is, for the upper quantiles of the CiC, the tangential shear is larger, given that a high galaxy number density should correspond to a large matter over-density on average. In order to extract cosmological information from this DSS, \citet{Friedrich:Gruen:2018} derived a lognormal model which predicts the shear profiles and the probability density of CiC by using the redshift distribution of sources, lenses, and the mean CiC as inputs. {In \citet{Gruen:Friedrich:2018}, the model was used to constrain cosmological parameters from DSS measurements from the Dark Energy Survey (DES) First Year and Sloan Digital Sky Survey (SDSS) data, where they included in their analysis the tangential shear profiles for scales greater than the top-hat filter size $\theta_\nt{th}$. Their analysis yields constraints on the matter density $\Omega_\nt{m}= 0.26^{+0.04}_{-0.03}$ that agree with the DES analysis of galaxy and shear two-point functions \citep[see][]{DSS:2018}.

\citet{Brouwer:2018} applied the DSS to the Kilo-Degree Survey \citep[KiDS;][]{Kuijken:2015} data, using the catalogue of
\citet{Bilicki:2018} for the foreground (lens) galaxies, whereas the source galaxies used for estimating the shear were taken from the third data release of KiDS \citep[see][]{deJong:2017}. In order to parameterised their measured shear signals, they fitted, for every quantile in the foreground galaxy CiC, a relation of the form $\gamma_{\nt{t}}=A/\sqrt{\theta}$ to their tangential shear profile, for $\theta>\theta_\nt{min}$, where $\theta_\nt{min}$ is approximately the radius of the peak of the shear profile.  By using this relation, they defined their signal-to-noise ratio $ \nt{S}/\nt{N}=A/\sigma_A$, where $\sigma_A$ is the 1$\sigma$ error on the best-fit amplitude based on the full analytical covariance matrix of the shear profiles. With a top-hat of size $5'$, they found for the regions with the highest 20$\%$ values of the aperture number a $\nt{S}/\nt{N}=21.7$ and for the lowest
20$\%$ a $\nt{S}/\nt{N}=16.9$. We use this fit relation later in this analysis to compare the S/N of our adapted filter to that of the top-hat filter.

The prime motivation for this work is based on the fact that the two components of the DSS -- the CiC of galaxies inside a radius $\theta$, and the tangential shear profile $\gamma_\nt{t}(\vartheta)$ for $\vartheta>\theta$ -- are poorly matched. For example, the shear at radius $\vartheta>\theta$ around a given point is affected by the matter distribution at all radii $<\vartheta$, not just by that inside $\theta$. Hence, even if the (foreground) galaxy density $n(\boldsymbol{\theta})$ had the same shape as the lensing convergence field $\kappa(\boldsymbol{\theta})$, the two
aforementioned quantities would not be perfectly correlated. Instead,
we consider here a pair of statistics for the foreground galaxy
distribution and the shear profile that are `matched', in the sense
that in the hypothetical case $n(\boldsymbol{\theta})\propto
\kappa(\boldsymbol{\theta})$, there would be a one-to-one relation
between them. This is achieved by using the aperture statistics \citep[see][]{Schneider:1996,Schneider:1998}, that is, aperture mass and aperture number counts. Although the case $n(\boldsymbol{\theta})\propto \kappa(\boldsymbol{\theta})$ is not a realistic assumption, due to different redshift weighting in the projected galaxy number density on the sky and the projected matter density between us and the lensed source population to obtain the convergence, we nevertheless expect a strong correspondence on the same angular scales, described by the galaxy-dark matter bias $b$ and correlation coefficient $r$ \citep{Pen:1998}. Instead of using the CiC, we now split the sky into areas of different quantiles of the aperture number counts, and consider the mean shear profile for each quantile; the latter is then quantified by the aperture mass. For the purpose of selecting a suitable filter function for the aperture statistics, we employ results from the ray-tracing through the Millennium Simulation \cite[hereafter MS;][]{Springel:2005,Hilbert:Hartlap:2009}, supplemented by a galaxy distribution obtained from a semi-analytic model \citep{Henriques:White:2015}. Hence, our filter function is adapted to expectations from cosmological simulations.

This work is structured as follows. In Sect.~\ref{Aperture_stat} we
review the basics of the aperture statistics. In Sect.~\ref{Data} we describe the simulation data used in this paper. Beside the MS, we use the Scinet Light Cone Simulations \citep[SLICS; see][hereafter HD18]{Harnois-Deraps:2018} to compare the performance of our new statistics to that of the previously employed DSS. For studying the sensitivity to cosmological parameters, we use the cosmo-SLICS simulations \citep[see][]{Harnois-Deraps:2019}, which are a suite of simulations for 26 different cosmologies. The derivation of the adapted filter is described in Sect.~\ref{newFilter}, and the comparison of the original DSS with our new method is performed in Sect.~\ref{determinationTophat}. In Sect.~\ref{MapNap_sec} we investigate the different relationships between the total matter and galaxy distribution for a non-linear and linear galaxy bias model. This is achieved by calculating aperture masses and aperture numbers with our new method and the method used in the previous DSS. In Sect.~\ref{CosCon} we compare both filters in their power to distinguish different cosmologies by use of cosmo-SLICS. In Sect.~\ref{Conclusion} we conclude and summarise our work. Furthermore, we give an outlook of possible future work and applications of our adapted filter.

%-------------------------------------------------------------------
\section{Aperture statistics}
\label{Aperture_stat}
Given a convergence (or dimensionless surface mass density) field
$\kappa(\boldsymbol{\theta})$, the aperture mass is defined as
\begin{equation}
    M _{\nt{ap}}\left(\boldsymbol{\theta}\right) \coloneqq \int\kappa(\boldsymbol{\theta}+\boldsymbol{\theta}')\,U(|\boldsymbol{\theta'}|)\,\dif^2\theta' \quad ,	
    \label{Map}
\end{equation}
where $ U(|\boldsymbol{\theta}|)$ is a compensated filter function,
such that $ \int\theta\, U(\theta)~ d\theta=0$. As shown in
\citet{Schneider:1996}, $M_{ \nt{ap}}$ can also be expressed in terms
of the tangential shear $\gamma_{\nt{t}}$ and a related filter
function $Q$ as
\begin{equation}
     M_{ \nt{ap}}(\boldsymbol{\theta}) = \int\,\gamma_{\nt{t}}(\boldsymbol{\theta}+\boldsymbol{\theta}')\,Q(|\boldsymbol{\theta}'|)\,\dif^2\theta' \quad ,
     \label{MapQ}
\end{equation}
where
\begin{equation}
 Q(\theta) = \frac{2}{\theta^2} \int\limits_0^{\theta}\theta'U(\theta')~  \dif\theta' - U(\theta) \quad ,
 \label{NewQ}
\end{equation}
which can be inverted, yielding
\begin{equation}
 U(\theta) = 2\int\limits_{\theta}^{\infty} \frac{Q(\theta')}{\theta'}~ \dif\theta' - Q(\theta) \quad .
\label{NewU}
\end{equation}
 
In analogy to $M_{ \nt{ap}}$, we define the aperture number counts 
\citep[][]{Schneider:1998}, or aperture number, as
\begin{equation}
    N_{ \nt{ap}}(\boldsymbol{\theta}) \coloneqq \int\,n(\boldsymbol{\theta}+\boldsymbol{\theta}')\,U(|\boldsymbol{\theta}'|)\, \dif^2\theta' \quad ,
    \label{Nap}
\end{equation}
where $ U(\boldsymbol{\theta})$ is the same filter function as in
Eq.~\eqref{Map} and $ n(\boldsymbol{\theta})$ is the galaxy number density on the sky. Our proposed modified DSS consists of splitting the sky into quantiles of $N_\nt{ap}$, and stacking the azimuthal-averaged tangential shear profile around all points of the given quantile. By setting $Q(\theta) = \gamma_\nt{t}(\theta)$, we then define a new $U$ filter for $N_\nt{ap}$ with Eq.~\eqref{NewU}, and iteratively repeat the process until we reach convergence (see Sect.~\ref{newFilter} for details). This differs from \citet{Gruen:2015} who determine the CiC from Eq.~\eqref{Nap} with a top-hat filter where
\begin{equation}
U_\nt{th}(\theta) = \mathcal{H}(\theta_\nt{th}-\theta) \quad,
\label{U_th_filter}
\end{equation}
with $\theta_\nt{th}$ is the size of the top-hat and $\mathcal{H}$ is the Heaviside step function. Since the top-hat filter $U_\nt{th}$ is not compensated,  we can not use Eq.~\eqref{NewQ} to calculate a corresponding filter
$Q_\nt{th}$. Instead, we set 
\begin{equation}
Q_\nt{th}(\theta) \sim  \begin{cases} 1/\sqrt{\theta} & \nt{, if } 1.2\,\theta_\nt{th}<\theta<\theta_\nt{max} \\0 & \,\nt{, otherwise} \end{cases} 
\label{Q_th_filter}
\end{equation}
following the work of \citet{Brouwer:2018}, where they used a $ 1/\sqrt{\theta}$ profile to parameterise their shear signals. The radius $\theta_\nt{max}$ is the size up to which we measure the shear profiles.

To efficiently calculate the aperture number we make use of the convolution theorem
\begin{equation}
N_{ \nt{ap}}(\boldsymbol{\theta}) = \mathcal{F}^{-1}\left[\mathcal{F}\{n(\boldsymbol{\theta})\}\mathcal{F}\{U(|\boldsymbol{\theta}|)\}\right] \quad ,
\label{conv_FFT}
\end{equation}
where $\mathcal{F}$ denotes the Fourier transformation and $\mathcal{F}^{-1}$ the inverse Fourier transformation \citep[see][hereafter
FFT]{Frigo:2005}.

\section{Mock KiDS data}

\label{Data}

In this work, we use three different simulation suites, which we modify to be KiDS-like. As KiDS is not so dissimilar from DES \citep[see][]{Drlica-Wagner:2018} and Hyper Suprime-Cam  \citep[see][]{Aihara:2019}, we expect our conclusion to also hold for these weak lensing surveys. We use the well tested MS to develop our adapted filter, and test our filter with an independent set of simulation, SLICS, to avoid recurring systematic effects. We also use the cosmo-SLICS to compare the adapted and top-hat DSS filters in their power to discriminate different cosmologies. 

\subsection{Millennium Simulation (MS)}
\label{MS_des}
The MS, described in \citet{Springel:2005}, follows the evolution of
$2160^3$ dark matter particles of mass $8\times10^8\,M_{\odot}/h$
enclosed in a cube of size $(500\,\nt{Mpc/}h)^3$. Galaxies are added
to the simulation afterwards using a semi-analytical galaxy-formation
model, where \citet{Saghiha:Simon:2017} showed that the best match
with the observed galaxy-galaxy lensing and galaxy-galaxy-galaxy
lensing signals from the Canada-France-Hawaii Telescope Lensing  Survey data \citep[see][]{Heymans:2012} 
is obtained from the model of \citet{Henriques:White:2015}. 
\citet{Hilbert:Hartlap:2009} described ray-tracing simulations 
through the MS. They constructed a suite of 64
pseudo-independent light cones of size $4\times4$\,deg$^2$. For each of them, they calculated the lensing Jacobi matrix $\mathcal{A}$ on a $4096\times4096$\,pixel grid, for a set of source redshifts, using a multiple lens plane algorithm. The Cartesian components of the shear for each grid point and each source redshift $z_\nt{c}$ are then obtained from the corresponding Jacobi matrix $\mathcal{A}$. We note that the same set of simulations has been used in several previous studies, for example, in \citet{Sadeh:Abdalla:2016}, \citet{Simon:Saghiha:2019}, and \citet{Unruh:2018,Unruh:2019}.

\subsubsection{Constructing foreground galaxy number densities}
To create the galaxy number density field $n(\boldsymbol{\theta})$ for each light cone, we project all galaxies with an SDSS
$r$-band magnitude $m_r<20.25$\,mag\footnote{These magnitudes are provided by the semi-analytical galaxy-formation
model.} onto pixels of size $(4\,\nt{deg}/4096)^2$. The magnitude cut is chosen such that the galaxy number density in the MS matches the one in \citet{Bilicki:2018}. The resulting redshift distribution of the galaxies over all 64 light cones is displayed in Fig.~\ref{redshiftdist} in orange, together with the redshift distribution of \citet{Bilicki:2018} shown in blue. We note that our lens redshift distributions is broader compared to \citet{Gruen:Friedrich:2018}; especially at small redshifts our lenses extend down to $z\approx0$.

\begin{figure}[!htbp]
\centering
\includegraphics[width=\linewidth]{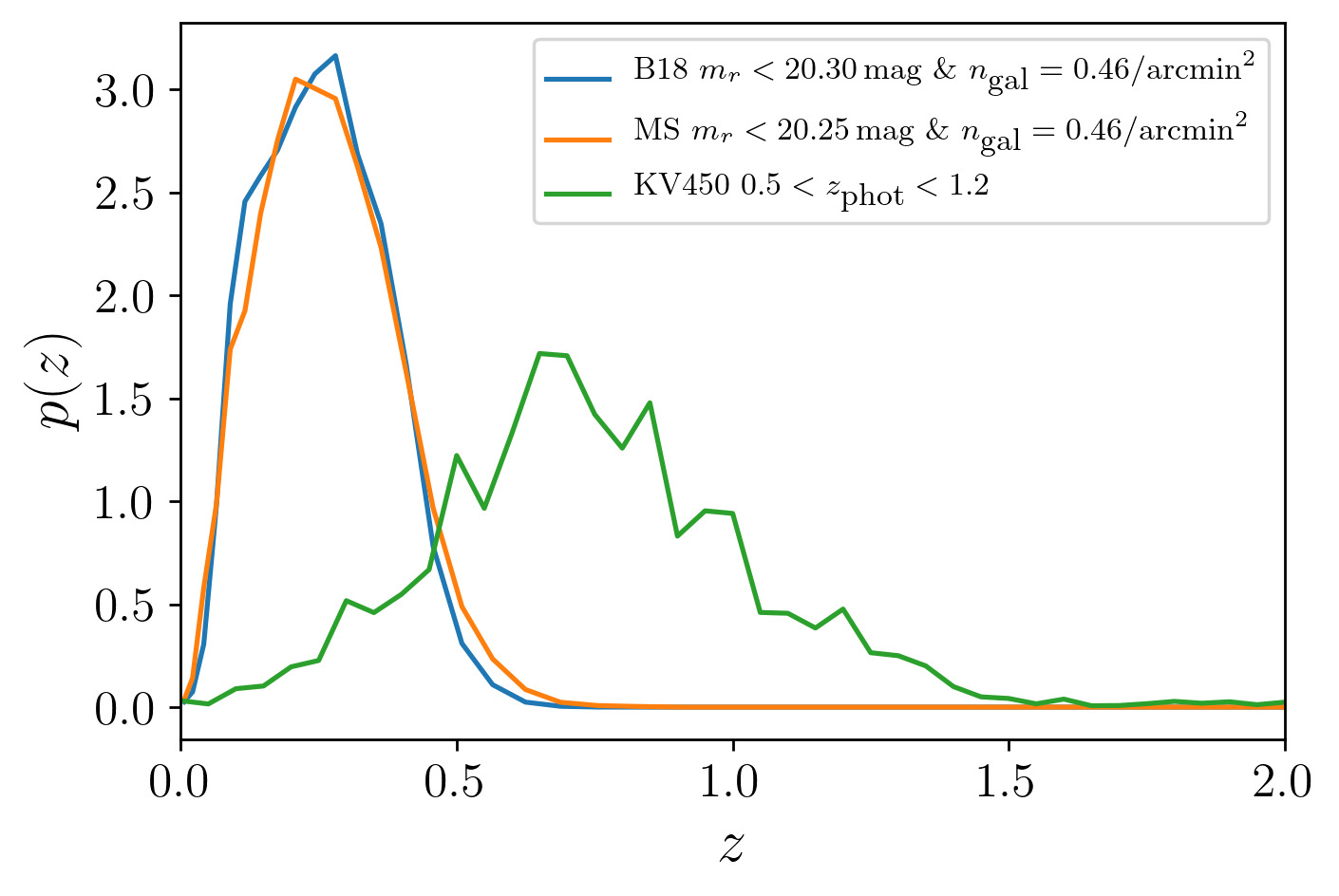}
\caption{Redshift distribution $p(z)$ of galaxies with $m_r<20.25$ in
  the 64 MS light cones, compared to the estimated redshift
  distribution of KiDS galaxies with $m_r<20.30$ \citep[][in the plot B18]{Bilicki:2018}. Shown in green is the weighted source redshift PDF of the highest three tomographic bins; from \citet{Hildebrandt:2018}.}
    \label{redshiftdist}
\end{figure}  

\subsubsection{Constructing the source galaxy distribution}
\label{MS_shear_cat}

In order to mimic the KiDS shear estimates, we create for each grid point in a light cone a weighted mean of the shear over all source
redshifts. We use the redshift distribution of the 
combined data set from the optical KiDS \citep[see][]{deJong:Kleijn:2013} and the near-infrared VISTA
Kilo degree Infrared Galaxy survey \citep[see][]{Edge:Sutherland:2013}. In this combined data set (hereafter KV450), redshifts are estimated through photometric redshifts, $z_\nt{phot}$, and calibrated with spectroscopic redshifts \citep[][]{Hildebrandt:2018}. We consider only sources with $0.5 < z_\nt{phot} < 1.2$, such that our sources are mostly behind our low-redshift lenses, and adopted the redshift distribution $n(z)$ from \citet[][]{Hildebrandt:2018} to model these sources (shown in green in Fig.~\ref{redshiftdist}). From this distribution the weights for the redshift slices $z_\nt{c}$ (see Sect.~\ref{MS_des}) of the simulation are calculated to
\begin{equation}
w(z_\nt{c}) = \int\limits_{z_{\nt{low}}(z_\nt{c})}^{z_{\nt{up}}(z_\nt{c})} p(z')\, \dif z' \quad ,
\end{equation} 
where $z_{\nt{low,up}}(z_\nt{c})$ are the boundaries of the
consecutive redshift slices in the MS with central redshift
$z_\nt{c}$.  With these weights, the shear at each grid point
$\gamma(\boldsymbol{\theta})$ is given as
\begin{equation}
\gamma(\boldsymbol{\theta}) = \frac{\sum\limits_i w(z_{\nt{c},i}) \, \gamma(\boldsymbol{\theta},z_{\nt{c},i}) }{\sum\limits_i w(z_{\nt{c},i})} \quad ,
\label{shear_MS}
\end{equation}
where $\gamma(\boldsymbol{\theta},z_{\nt{c},i})$ is the shear value at position $\boldsymbol{\theta}$ from the $i$-th redshift slice  calculated with the corresponding Jacobi matrix $\mathcal{A}$. Since the MS are exclusively used to construct our new filters, it is best to ignore shape noise, hence we work directly with the noise-free shear values provided with Eq.~\eqref{shear_MS}.

\subsection{Scinet Light Cone Simulations (SLICS)}
\label{SLICS}
In order to compare the performance of our adapted filter to that of the \citet{Gruen:2015} top-hat filter, and to find the appropriate size of the top-hat filter such that the comparison is reasonable, we use the SLICS. This simulation suite is independent of the MS and is described in HD18. The SLICS are a set of over 800 realisations, where each run follows 1536$^3$ particles inside a  cube of comoving side length $L_{\nt{box}}=505\,h^{-1}\nt{Mpc}$ and $n_\nt{c}=3072$ grid cells on the side. By use of the Zel'dovich approximation \citep[see][]{White:2014} each run starts with slightly different initial conditions at $z=120$, computes the non-linear evolution of these collision-less particles to $z=0$,
and produces on-the-fly the halo catalogues and mass sheets required for a full light cone construction at 18 different source redshifts from $z=0$ to $z=3$. The underlying cosmological parameters for each run are $\Omega_{\nt{m}}=0.2905$, $\Omega_{\Lambda}=0.7095$,
$\Omega_{\nt{b}}=0.0473$, $h=0.6898$, $\sigma_8=0.826$ and
$n_s=0.969$ \citep[see][]{Hinshaw:2012}. Given a particle mass of $2.88\times10^{9}\,h^{-1}\nt{M}_{\odot}$, dark matter haloes with masses above $10^{11}\,h^{-1}\nt{M}_{\odot}$ and structure formation deep into the non-linear regime are resolved. Furthermore, it has been shown in HD18 that for Fourier modes $k<2.0\,h\,\nt{Mpc}^{-1}$, the three-dimensional
dark matter power spectrum $P(k)$ agrees within 2$\%$ with the
predictions from the Extended Cosmic Emulator \citep[see][]{Heitmann:Lawrence:2014}, followed by a progressive
deviation for higher $k$-modes.

\subsubsection{KV450 SLICS mocks}
We use the KV450 SLICS as source galaxies\footnote{These SLICS KV450
mocks are made publicly available on the SLICS portal at
\url{https://slics.roe.ac.uk/}}. These
mock galaxies are placed at random angular coordinates on 100\,deg$^2$ light cones, with the KV450 number density
$n_{\nt{gal}}=6.93/$arcmin$^2$ and the best-estimated redshift
distributions from \citet[][see the DIR method
therein]{Hildebrandt:2018}. The galaxies are assigned their shear information $\gamma$ from the lensing maps, following the linear interpolation algorithm described in Sect. 2 in HD18; and the observed ellipticities $\epsilon^{\nt{obs}}$ are obtained as
\begin{equation}
\epsilon^{\nt{obs}} = \frac{\epsilon^{\nt{int}}+\gamma}{1+\epsilon^{\nt{int}}\gamma^{*}} + \eta \approx \frac{\epsilon^{\nt{n}}+\gamma}{1+\epsilon^{\nt{n}}\gamma^{*}} \quad ,
\end{equation} 
where $\epsilon^{\nt{obs}}$, $\epsilon^{\nt{int}}$,
$\epsilon^{\nt{n}}$, $\eta$, and $\gamma$ are complex numbers; the
asterisk $*$ indicates complex conjugation. This equation relates the
observed ellipticity $\epsilon^{\nt{obs}}$ to the intrinsic shape
$\epsilon^{\nt{int}}$ and the shear $\gamma$, and adds measurement
noise $\eta$ to it. In order to combine intrinsic and measurement shape noise, both are incorporated into one pre-sheared noisy ellipticity $\epsilon^{\nt{n}}$. This ellipticity $\epsilon^{\nt{n}}$ is generated by drawing random numbers from a Gaussian distribution with width $\sigma=0.29$, which is consistent with the weighted observed ellipticity distribution of the KiDS data. Furthermore, we apply a selection cut on the photometric redshift of $0.5<z_{\nt{phot}}<1.2$, resulting in a galaxy number density of $n_{\nt{gal}}=5.17/$arcmin$^2$.

\subsubsection{Galaxy And Mass Assembly (GAMA) SLICS mocks}
\label{GAMA_mocks}
For the lens sample we use the publicly available Galaxy And Mass Assembly \citep[GAMA, see][]{Driver:2011} SLICS mocks, which are based on the halo occupation distribution (HOD) prescription of \citet[][see HD18 for details on its
implementation]{Smith:2017}. The motivation to use these mocks is that they are an excellent source of lenses for a DSS analysis with KiDS data as sources, as demonstrated by \citet{Brouwer:2018}. The galaxy number density is $n_{\nt{gal}}\sim0.25/$arcmin$^2$, which is smaller compared to \citet{Bilicki:2018} due to the smaller limiting magnitude of $m_r<19.8$\,mag and the smaller redshift range of $0<z<0.5$, but since we use both data sets in two independent analyses it does not matter. Theses different values of $n_{\nt{gal}}$ propagate into the aperture number, $N_\nt{ap}$ via Eq.~\eqref{Nap}, where we count these GAMA lens galaxies in squares of size 1\,arcmin$^2$ and assign the resulting galaxy number density $n(\boldsymbol{\theta})$ to the associated pixel. Finally, it was demonstrated in HD18 that on large scales these mock GAMA galaxies have a linear bias of about 1.2, and that the non-linear bias observed at smaller scales is similar to that seen in the GAMA data. This match was not guaranteed given that the galaxy bias in the simulations emerge from the HOD, and not from an input model.

\subsection{Cosmo-SLICS}
\label{Cosmo-SLICS}

We use the cosmo-SLICS simulations described in \citet{Harnois-Deraps:2019}, to investigate the sensitivity of the top-hat filter and the adapted filter to cosmological parameters. These are a suite of simulations sampling 26 $w$CDM cosmologies distributed in a Latin hypercube, ray-traced multiple times to produce 50 pseudo-independent realisations for every cosmology, each producing
light cones of size 100\,deg$^2$. The corresponding cosmologies are listed in Table~\ref{cos_overview}. In these simulations, the matter
density $\Omega_{\nt{m}}$, the dimensionless Hubble parameter $h$, the
normalisation of the matter power spectrum $\sigma_8$ and the time-independent dark energy
equation-of-state $w_0$ are varied over a range that is large enough to complement the analysis of current weak lensing data \citep[see][]{Hildebrandt:2018}.

For each realisation, the algorithm to creates KV450-like catalogues follows the same pipeline as for the SLICS mocks, notably it reproduces the same galaxy number density and redshift distribution $n(z)$, but the different underlying cosmologies modify the lensing properties.

In contrast to the SLICS simulations, the cosmo-SLICS dark matter haloes are not fully post-processed into light cones at the moment of writing this paper, and therefore HOD-based mocks are not yet available. This does not prevent us from using the cosmo-SLICS to generate GAMA-like mocks, however these are instead based on a linear bias model (see Sect. A2 of HD18). Given the GAMA $n(z)$, this construction required four mass sheets\footnote{For the fiducial cosmology these mass sheets are at redshifts $z_i = 0.130, 0.221, 0.317, 0.410$.}. Following the redshift distribution shown in Fig.\,8 in HD18 each of these sheets was populated with a bias of unity, and  accordingly to Sect.~\ref{GAMA_mocks} the resulting number density for all four sheets together is $n_{\nt{gal}}=0.25/$arcmin$^2$. To be consistent with Sect.~\ref{GAMA_mocks}, we sum the galaxies in squares of size 1\,arcmin$^2$ and assign the galaxy number density $n(\boldsymbol{\theta})$ to the
respective pixels.

\section{The derivation of the adapted filter function}
\label{newFilter}
In order to investigate the projected galaxy number density
$n({\boldsymbol\theta})$ and lensing convergence $\kappa(\boldsymbol\theta)$ on the same
angular scales, we generate a compensated filter for $\theta <
30'$ using an iterative procedure with the MS as an input. Schematically the iterative process is structured as follows: The first step is to calculate the aperture number $N_{\nt{ap}}$ with a compensated filter $U_i$ defined for $\theta < 10'$. Next, we extract the pixels which have the highest 10$\%$ aperture number values, and measure the tangential shear profile $\gamma_\nt{t}(\theta)$ around these pixels up to $10'$. With setting $Q(\theta)\propto\gamma_\nt{t}(\theta)$ and Eq.\,(\ref{NewU}) we create a revised compensated filter $U_{i+1}$. The last step is to repeat all prevoius steps with the revised filter $U_{i+1}$. This iteration continues as long as the change in relative signal-to-noise $\Delta \nt{(S/N)}/\nt{(S/N)}_1 > 10^{-3}$ between consecutive iterations. We note that this value is chosen arbitrarily, but it is sufficient, because the deviation of the resulting shear profiles in Sect.~\ref{determinationTophat}, determined with a filter of a later iteration, would be less than the uncertainties of the shear profiles. Once we achieve convergence in this iterative process, we extrapolate the $U$ and $Q$ filters to $30'$ to use the strong tangential shear signal beyond $10'$.

After presenting the general approach of our derivation, we next
explain the individual steps in more detail. The initial filter $U_1$
of the pipeline is defined as a compensated top-hat
\begin{equation}
    U_1(|\boldsymbol{\theta}|)\coloneqq \begin{cases}1\,\nt{arcmin}^{-2} & \nt{, if } \theta < 1'\\-\frac{1}{99} \,\nt{arcmin}^{-2} & \nt{, if } 1' \leq \theta < 10' \\ 0 \,\nt{arcmin}^{-2} & \nt{, if } \theta > 10' \end{cases} \quad ,
    \label{StartingU}
\end{equation}
where the chosen inner radius of $1'$ is not crucial, because the iterative process finds the final shape of the
filter independent of this boundary. The upper bound of $10'$ is motivated by the fact that we expect the shear profiles with our filter to peak at roughly $2/3$ of the filter size, which would then coincide with the shear profiles generated with a top-hat filter of size $5'$ in \citet{Brouwer:2018} which had the best S/N. The value $-1/99\,\nt{arcmin}^{-2}$ arises from the compensated nature of $U$. To calculate the aperture number with Eq.~\eqref{Nap}, we convolve the galaxy number density $
n(\boldsymbol{\theta})$ with the filter $U_1$ by means of the
convolution theorem Eq.~\eqref{conv_FFT}. The resulting aperture number for one light cone is shown in the upper panel of Fig.~\ref{NapFigure},
where over-dense regions are shown in red and under-dense regions in
blue. Following the pipeline, we extract those pixels that
have the highest $10\%$ values of the aperture number and display them in the
lower panel of Fig.~\ref{NapFigure}. The outer 30$'$ edges are not
considered since the FFT, which we use to efficiently apply the
convolution theorem, assumes periodic boundary conditions. The reason to cut at 30$'$ instead of 10$'$ is that we want to use the same area of the light cones for the extended shear profile as for the ones measured in the iterative process.

\begin{figure}[!htbp]
\begin{minipage}{\linewidth}
    \includegraphics[width=\linewidth]{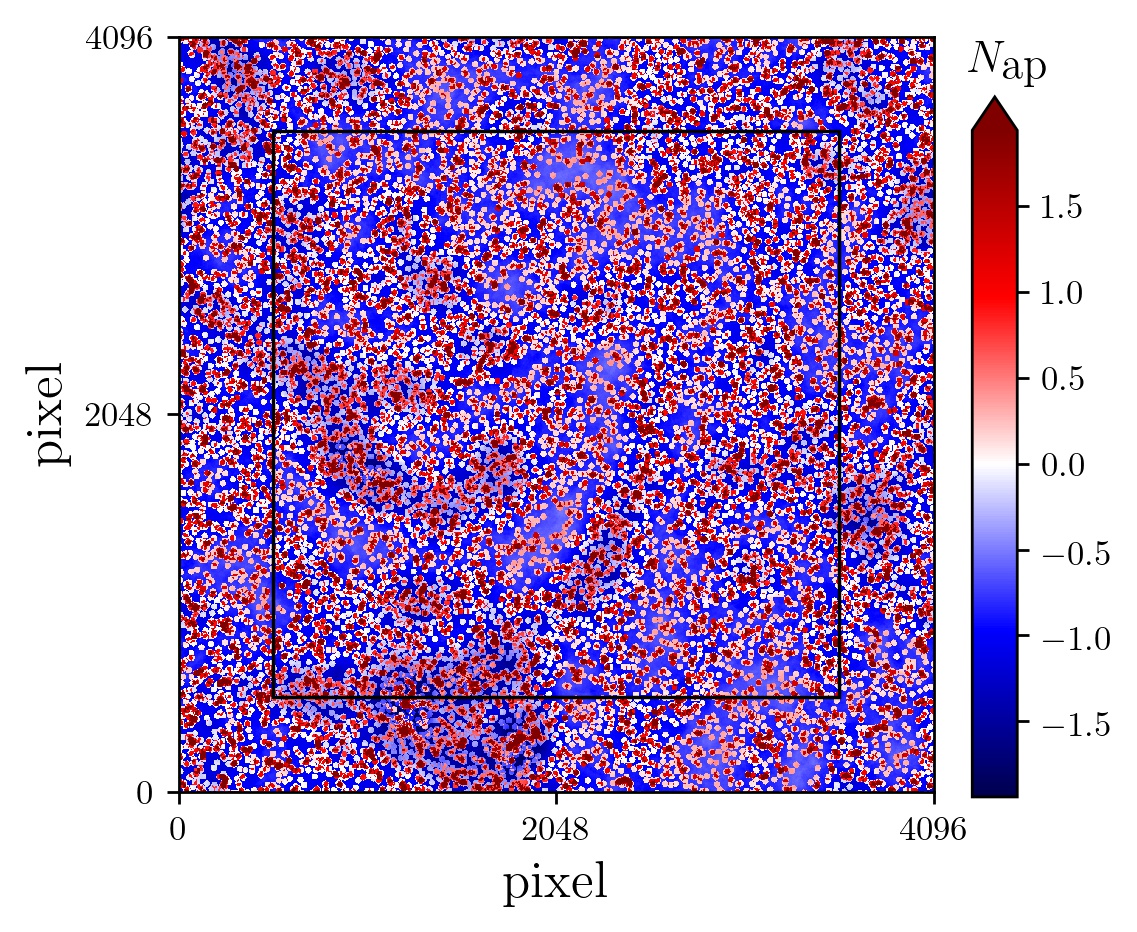}
\end{minipage}
\begin{minipage}{\linewidth}
    \includegraphics[width=\linewidth]{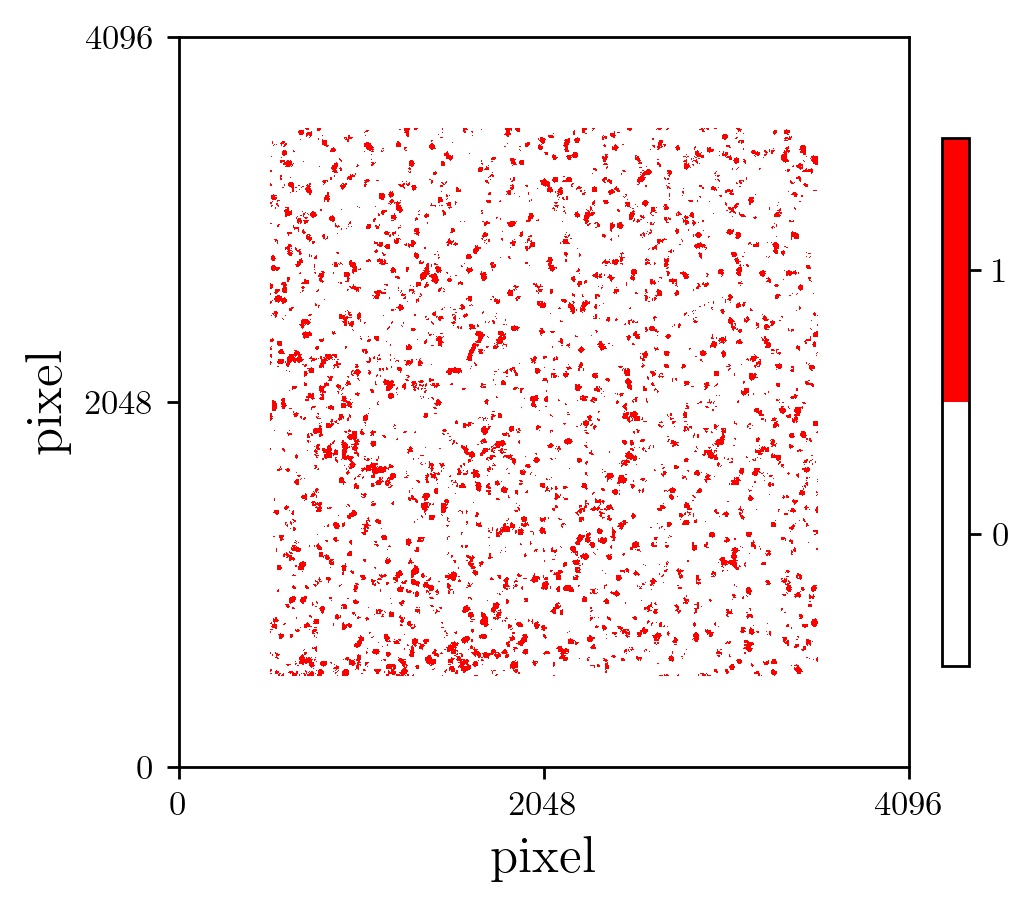}
\end{minipage}
\caption{Upper panel: Aperture number $N_{\nt{ap}}$ on a
  $4\times4\,\nt{deg}^{2}$ grid of the MS light cone 37 as an example light cone. Lower panel: Extracted pixels which have the highest 10$\%$  number values of $N_{\nt{ap}}$. The outer $30'$ margins are not considered since the FFT assumes periodic boundary conditions. Therefore, the outer margins in the $ N_{\nt{ap}}$ field are disregarded and marked with the black square in the upper panel.}
    \label{NapFigure}
\end{figure}

Using the shear grids described in Sect.~\ref{MS_des}, an averaged shear grid around the extracted pixels is calculated as 
\begin{equation}
    \overline{\gamma}(\boldsymbol{\theta})= \frac{1}{N_\nt{peaks}} \sum \limits_{i=1}^{N_\nt{peaks}} \gamma(\boldsymbol{\theta}+\boldsymbol{\theta}_i)\quad ,
\label{gamma_stack}
\end{equation}
where $N_\nt{peaks}$ is the number of extracted pixels with positions
$\boldsymbol{\theta}_i$, which have the $10\%$ highest values of $
N_{\nt{ap}}$. Next, we construct the grids of tangential and cross shear
$\gamma_{{\nt{t}},\times}(\boldsymbol{\theta})$, with
\begin{equation}
  \gamma_{\nt{t}}(\boldsymbol{\theta}) = 
- \mathfrak{Re}\left[\overline{\gamma}(\boldsymbol{\theta})
\nt{e}^{-2\nt{i}\phi}\right]  
\quad ; \quad 
\gamma_{\times}(\boldsymbol{\theta}) = 
- \mathfrak{Im}\left[\overline{\gamma}(\boldsymbol{\theta}) 
\nt{e}^{-2\nt{i}\phi}\right] \quad ,
\end{equation}
where $\phi$ is the polar angle of $\boldsymbol{\theta}$. For all shear profiles we subtract the shear signal around random pixel positions per light cone to reduce the noise in the measurements \citep[][]{Singh:Mandelbaum:2017}.

The shear profiles for one light cone result from azimuthally
averaging the $\gamma_{{\nt{t}},\times}(\boldsymbol{\theta})$ grids in
40 linearly spaced annuli for $0’<\theta<10’$. By further averaging the signals over all 64 light cones, we extract the shear profiles indicated with the blue dots in Fig.~\ref{gamma_profile}, where the error bars are the uncertainties on the mean, obtained from the sample variance of all 64 light cones. In the lower panel the $\gamma_{\times}$ profile is displayed, and although a $40\times40$ covariance cannot be reliably calculated from only 64 realisations, the cross shear profiles appear to be consistent with zero. The shape of the $\gamma_{\nt{t}}$ profiles are as expected for a DSS
analysis and similar to those of previous DSS works \citep{Brouwer:2018,Gruen:Friedrich:2018,Friedrich:Gruen:2018}.

\begin{figure*}[!htbp]
%\centering
%\sidecaption
    \includegraphics*[width=\linewidth]{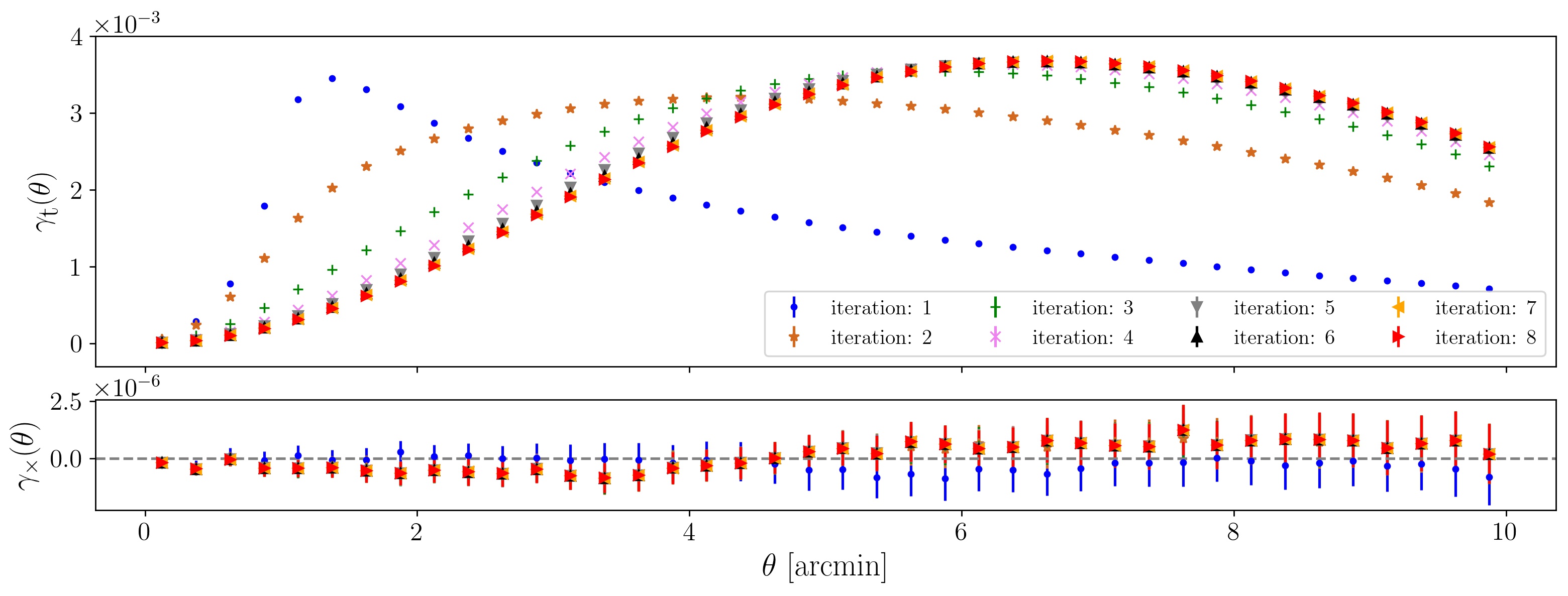}
    \caption{Upper panel: Tangential shear profiles, $\gamma_\nt{t}$, for the first eight iterations, showing how the peak moves to larger radii. Lower panel: $\gamma_{\times}$ profiles are consistent with zero. The uncertainties are the standard deviation on the mean determined with the 64 MS realisations.}
    \label{gamma_profile}
\end{figure*}
\begin{figure*}[!htbp]
%\sidecaption
\includegraphics*[width=\linewidth]{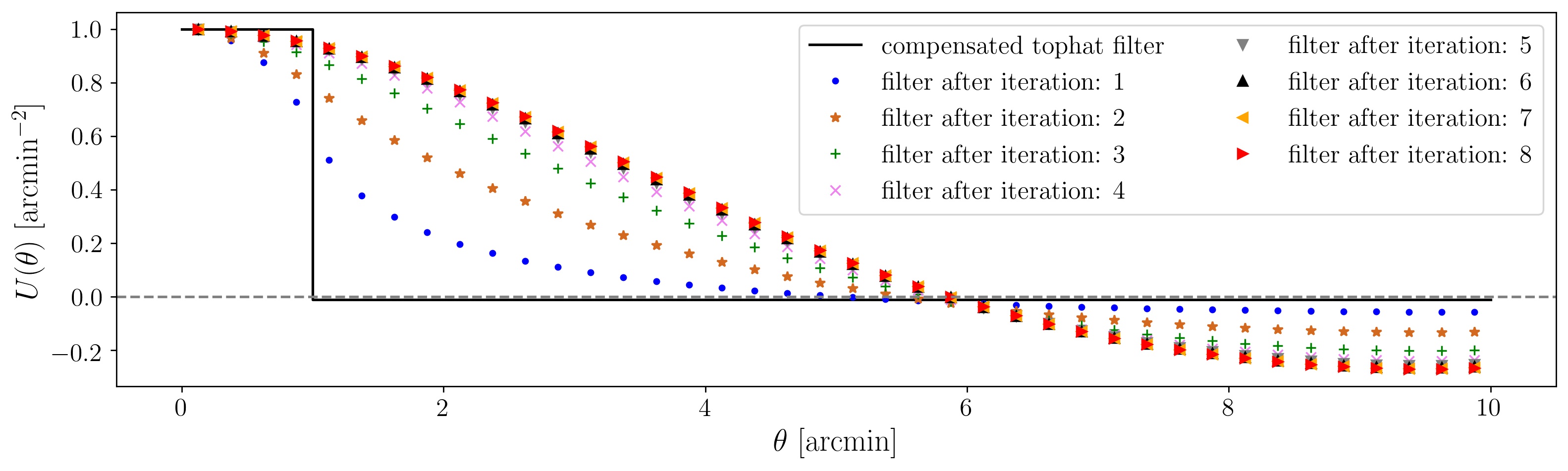}
\caption{Resulting filter $U$ , from Eq.~\eqref{NewU}, after each iteration. With each iteration the filter gets wider until it converges after $\sim 7$ iterations. The filters are scaled such that the value of the first $\theta$-bin is unity, which eases comparison with the compensated top-hat filter.}
    \label{NewFilterU}
\end{figure*}
\begin{figure}[!htbp]
\centering
    \includegraphics[width=\linewidth]{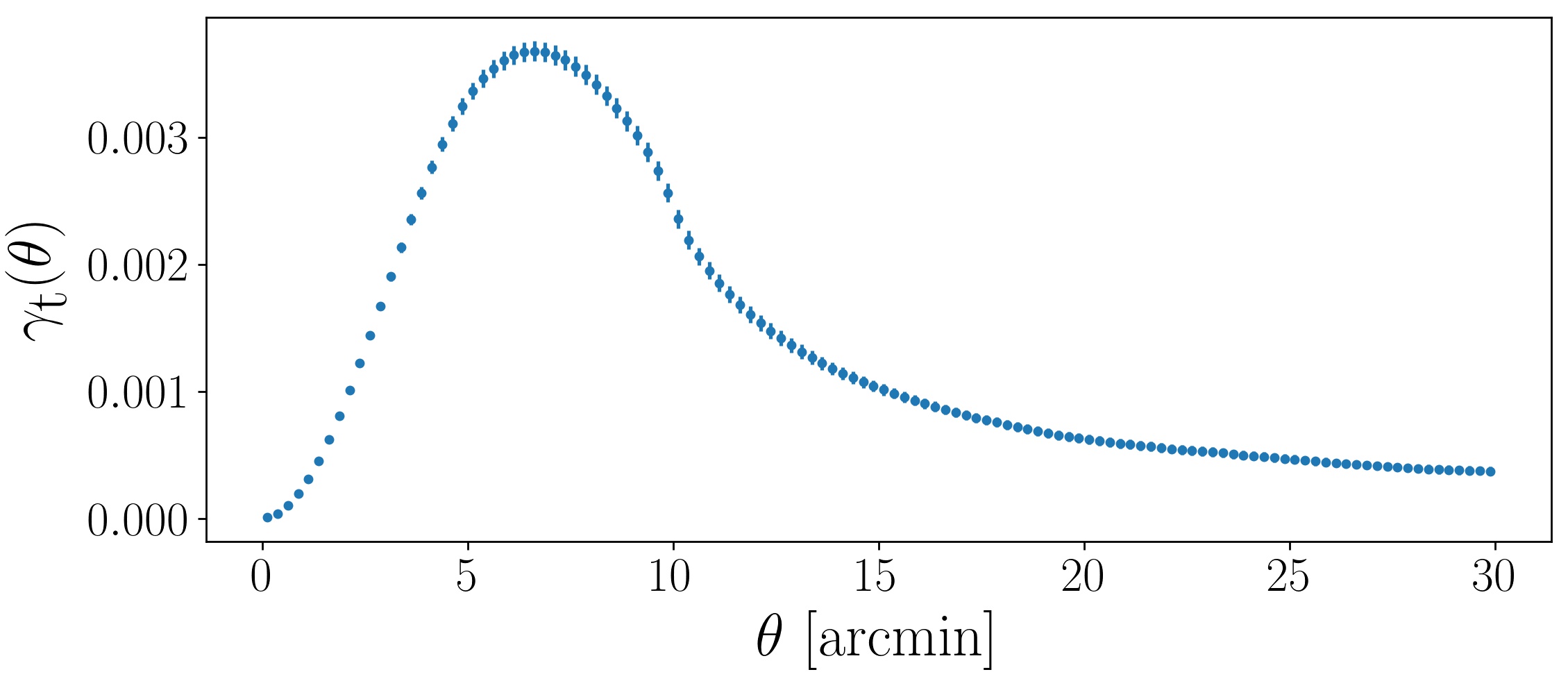}
    \caption{Tangential shear profile, $\gamma_\textrm{t}$, around the highest 10$\%$ pixel values of $N_\nt{ap}$ determined with the filter $U_7$ for $\theta <10'$ and measured up to radii of $30'$ to use the strong tangential shear signal beyond $10'$. For the rest of this analysis, this is the shape of the adapted filter $Q$. The uncertainties are the standard deviation on the mean determined with the 64 MS realisations. }
    \label{Final_Filter_Q}
\end{figure}
\begin{figure}[!htbp]
\centering
    \includegraphics[width=\linewidth]{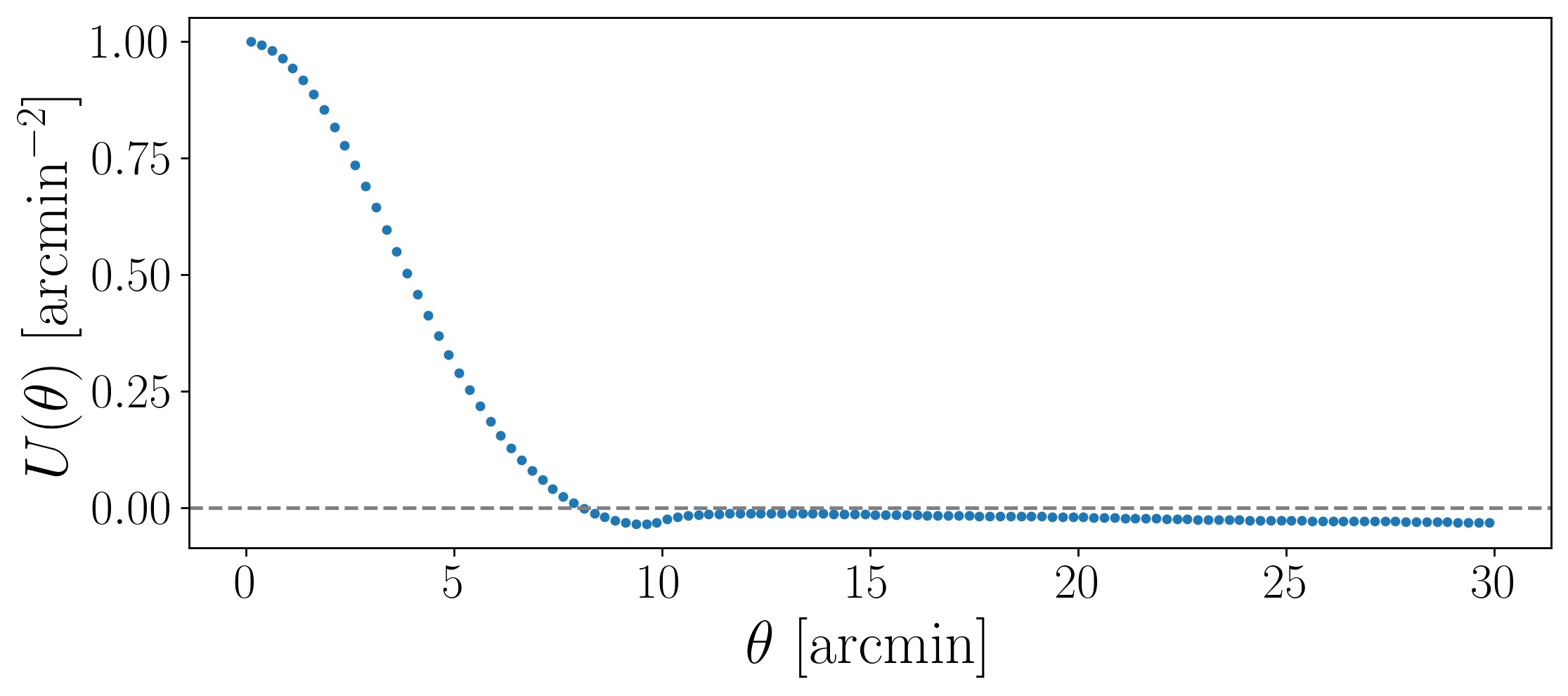}
    \caption{Adapted compensated filter $U(\theta)$ calculated from
      the shear profile of Fig.~\ref{Final_Filter_Q} and
      Eq.~\eqref{NewU}. The filter is normalised such that the first
      value is 1\,arcmin$^{-2}$. This final $U(\theta)$ filter is adopted for the rest of the analysis.}
    \label{Final_Filter_U}
\end{figure}

For determining the filter function $U$, we quantify the information content about these shear profiles through  $M_\nt{ap}$, by defining a signal-to-noise ratio \citep[][]{Schneider:1996}, 
\begin{equation}
    \frac{\nt{S}}{\nt{N}} = \frac{\sqrt{2}}{\sigma_\epsilon} \frac{\sum_{i} \gamma_{\nt{t}}(\theta_i)\,Q(\theta_i)}{\sqrt{\sum_{i} Q^2(\theta_i)}} \quad , 
   \label{S/N}
\end{equation}
where the noise here is taken to be pure shape noise 
due to intrinsic source ellipticity, with a dispersion of $\sigma_\epsilon=0.3$. 

The next step of the pipeline is motivated by Eq.~\eqref{S/N}, which following the Cauchy–Schwarz inequality, is maximised if the filter $Q$ is proportional to the shear $\gamma_{\nt{t}}$. We therefore set the $Q_2$ filter function to
\begin{equation}
    Q_2(\theta) \propto \begin{cases}\gamma_{\nt{t}}(\theta) & \nt{, if } 0' < \theta < 10'\\ 0 & \nt{, otherwise} \end{cases}\quad .
    \label{Q_propto}
\end{equation}
With this filter $Q_2$ and Eq.~\eqref{NewU}, we obtain filter $U_{2}$, displayed with the blue colour in Fig.~\ref{NewFilterU}.

Now the iterative process starts, where we rerun the pipeline with
the new filters $U_{i+1}$. As seen in Fig.~\ref{gamma_profile} the
peak of the tangential shear moves to larger radii after each
iteration, as the filter $U_{i+1}$ gets wider with each
iteration. This effect is not surprising, because we are calculating the filters $U_{i+1}$ from the shear signals, and therefore, the changes are strongly related. After some iterations this broadening starts to converge; in order to measure this convergence, we make use of the S/N calculated with Eq.~\eqref{S/N}. As a reference S/N value for the first iteration, we calculate an initial filter $Q_1$ from Eq.~\eqref{NewQ} as
\begin{equation}
Q_1(\theta) =  \frac{1}{\theta^2} \left(1+\frac{1}{99}\right) \mathcal{H}(\theta-1')\mathcal{H}(10'-\theta) \quad .
\end{equation}
The resulting S/N values
relative to the S/N of the first iteration are stated in
Table~\ref{tableS2N}. The S/N does not change
after the 7th iteration by more than $10^{-3}$, and therefore indicates convergence.
\begin{table}[!htbp]
\caption{S/N relative to the S/N of the first iteration step}
\centering
\begin{tabular}{c|cccccccc}
step & 1 & 2 & 3 & 4    \\
\hline
(S/N)/(S/N)$_1$ & 1. & 1.819 & 2.189 & 2.257  \\ 
\hline
\hline
step & 5 & 6 & 7 & 8   \\
\hline
(S/N)/(S/N)$_1$ & 2.271 & 2.274 & 2.275 & 2.275  \\ 
\end{tabular}
\label{tableS2N}
\end{table}

Once we	have converged on a final filter $U_7$, we expand the range up to a radius of $30'$ to make use of the strong tangential shear signal beyond $10'$. This size is restricted to $30'$ to minimise the rejected margins due to the boundary effects of the FFT, as seen in the lower panel of Fig.~\ref{NapFigure}. The resulting shear profile, and thus the shape of the final adapted filter $Q$, is shown in Fig.~\ref{Final_Filter_Q}. The corresponding adapted filter $U$, from Eq.~\eqref{NewU} using the extended adapted filter $Q$, is displayed in Fig.~\ref{Final_Filter_U}. Compared to the filters in Fig.~\ref{NewFilterU} the zero crossing of the
adapted filter $U$ is at larger $\theta$. This is due to the positive extended tail of the tangential shear profile (adapted filter $Q$ in Fig.~\ref{Final_Filter_Q}), which is used to determine the adapted filter $U$. After this point we do not change this filter anymore, and all filter functions mentioned from now on refer to this pair of adapted filters. We note that the used angular scales for the derivation of the filter function (1$'$,10$'$ and 30$'$) may not be optimal, but for the purpose of this work in comparing it to an analysis with a top-hat filter function (Sect.~\ref{determinationTophat},~\ref{MapNap_sec},~\ref{CosCon}) the optimised sizes are not crucial. Nevertheless, they will be reviewed in future analyses. Since the comparisons of the adapted and the top-hat filter in the following sections is exclusively done with the SLICS and cosmo-SLICS, the MS is from this point on no longer used. 

\section{Suitable top-hat size and S/N comparison}
\label{determinationTophat}

In order to compare the DSS measured using the adapted filter $U$ to the \citet{Gruen:2015} top-hat filter function $ U_\nt{th}(\theta)$
we must determine the size of the top-hat filter $\theta_\nt{th}$ such that the averaged shear peak positions around the highest and the lowest quantiles of the aperture number field are comparable between both filters. For that, we use 64 realisations of SLICS, with KV450 sources and GAMA lenses. 
 
Following the work of \citet{Gruen:Friedrich:2018}, we divide the sky
according to the aperture number, $N_\nt{ap}$, into five sub-areas of equal size and call them quantiles of the aperture number field. The aperture number is calculated either with the adapted filter function $U$ or with three different top-hat filters of size $\theta_\nt{th}=5'$, $6'$, and $7'$. For each quantile, we calculate the tangential shear profiles in 25 logarithmic $\theta$ annuli with the software \textsc{treecorr} \citep[see][]{Jarvis:Bernstein:2004}; this is different to the approach in Sect.~\ref{newFilter}, since for the SLICS and cosmo-SLICS the shear estimates are not given on a grid but from mock catalogues. The resulting shear profiles are displayed for the different filter functions in Fig.~\ref{shear_comp}. We neglect shape noise here to find the optimal top-hat size.

\begin{figure*}[!htbp]
%\sidecaption
    \includegraphics*[width=\linewidth]{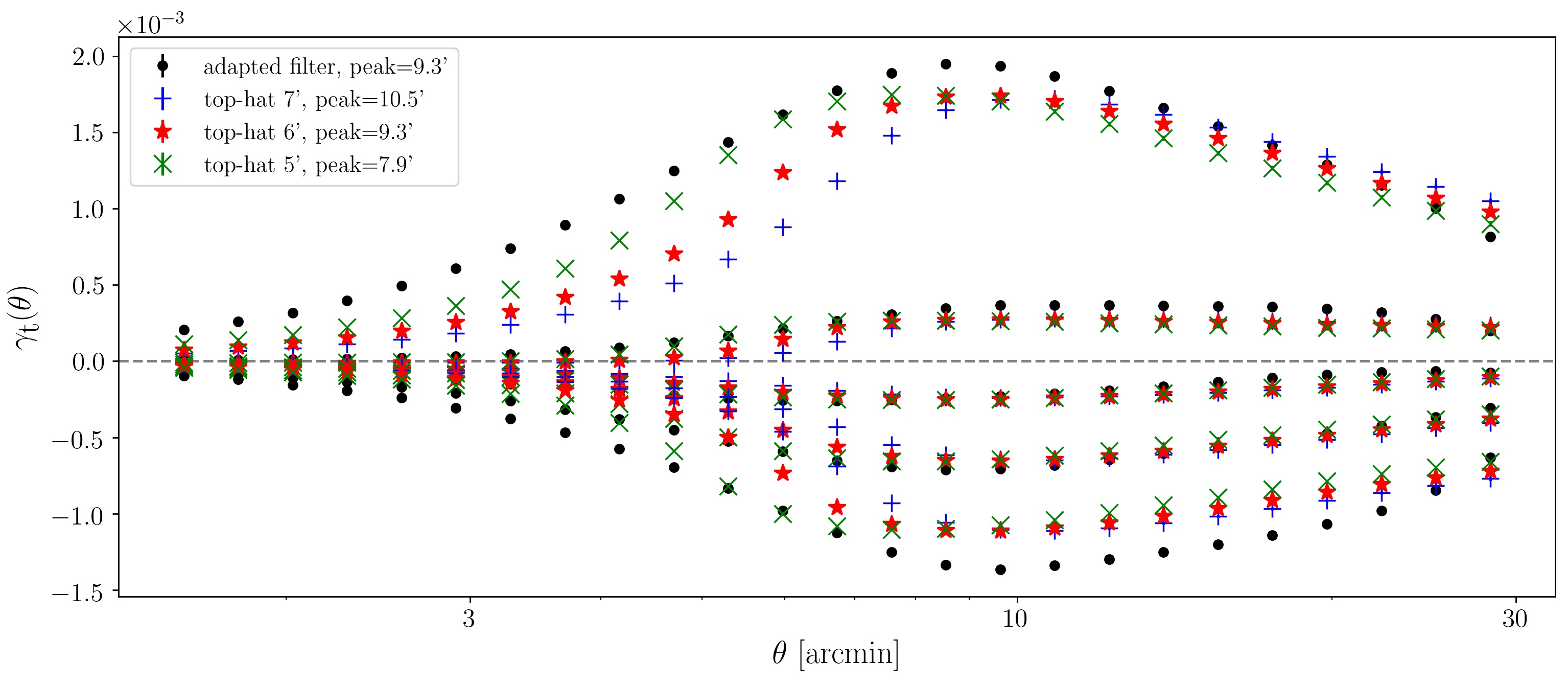}
    \caption{Tangential shear profiles, $\gamma_\textrm{t}$, from SLICS generated with the
      adapted filter and three top-hat filters of different
      sizes. The measurements using a top-hat of size $6'$ have roughly the same peak position as the adapted filter results. The uncertainties are the standard deviation on the mean determined with the 64 SLICS realisations, and since shape noise is not included, the error bars are almost unseen.}
    \label{shear_comp}
\end{figure*}

In order to determine the most comparable top-hat filter, we calculate for each filter the angular position of the measured peak of the $\gamma_\nt{t}$ profile of the highest and lowest quantile and report in the legend of Fig.~\ref{shear_comp} the average of these two. The averaged peak position $\theta = 9.3'$ of the shear profiles generated with a top-hat filter of size $\theta_\nt{th}=6'$ matches the averaged peak position $\theta=9.3'$ of the shear profiles generated with the adapted filter. Therefore, we set the size of the top-hat filter for all following analyses to $\theta_\nt{th}=6'$.

Our first performance comparison is based on a respective S/N. The signal S is the averaged aperture mass for axis-symmetric tangential shear profiles $\gamma_{\nt{t}}(\boldsymbol{\theta}) = \gamma_{\nt{t}}(\theta)$, such that Eq.~\eqref{MapQ} simplifies to
\begin{equation}
     M_{ \nt{ap}}^i = 2 \pi \int  \gamma_{\nt{t}}^i(\theta')\,Q(\theta')\,\theta'\dif \theta' \quad ,
     \label{MapQ_polar}
\end{equation}
where $i$ denotes the quantile around which the tangential shear profile $\gamma_{\nt{t}}^i(\theta)$ is azimuthal-averaged. To calculate the aperture mass with the tangential shear profiles of the DSS with the top-hat filter, we use Eq.~\eqref{Q_th_filter} for the $Q=Q_\nt{th}$ filter with $\theta_\nt{max}=30'$. We reiterate that $Q_\nt{th}$ is not adapted to $U_\nt{th}$, but we use it here to provide a comparison to the earlier work of \citet{Brouwer:2018}. In order to have a S/N, which measures the significance of a nonzero detection, we estimate the noise N as the standard deviation of $M_\nt{ap}^\nt{rand}$ determined by tangential shear profiles around $N_\nt{pix}$ random pixel positions from the 64 realisations, where $N_\nt{pix}$ is the number of pixels in one quantile. Together this gives the signal-to-noise ratio of the $i$-quantile to
\begin{equation}
\left(\frac{\nt{S}}{\nt{N}}\right)^i = \frac{\langle M_{ \nt{ap}}^i \rangle}{\sqrt{\langle (M_\nt{ap}^\nt{rand}-\langle M_\nt{ap}^\nt{rand}\rangle)^2\rangle}} \quad,
\label{S/N_i}
\end{equation}
where $\langle...\rangle$ refers to the ensemble average over all 64 realisations. For this S/N comparison we use the \textsc{treecorr} $\gamma_{\rm t}$ estimates obtained from ellipticities with shapes noise, so that the noise here describes the sampling variance as well as the shape noise in the data. The resulting S/N for each quantile $i$, shown in Fig.~\ref{SLICS_S2N}, reveals that the adapted filter performs better, which is consistent with the higher amplitude of the shear profiles seen in Fig.~\ref{shear_comp}.  
\begin{figure}[!htbp]
\centering
    \includegraphics[width=\linewidth]{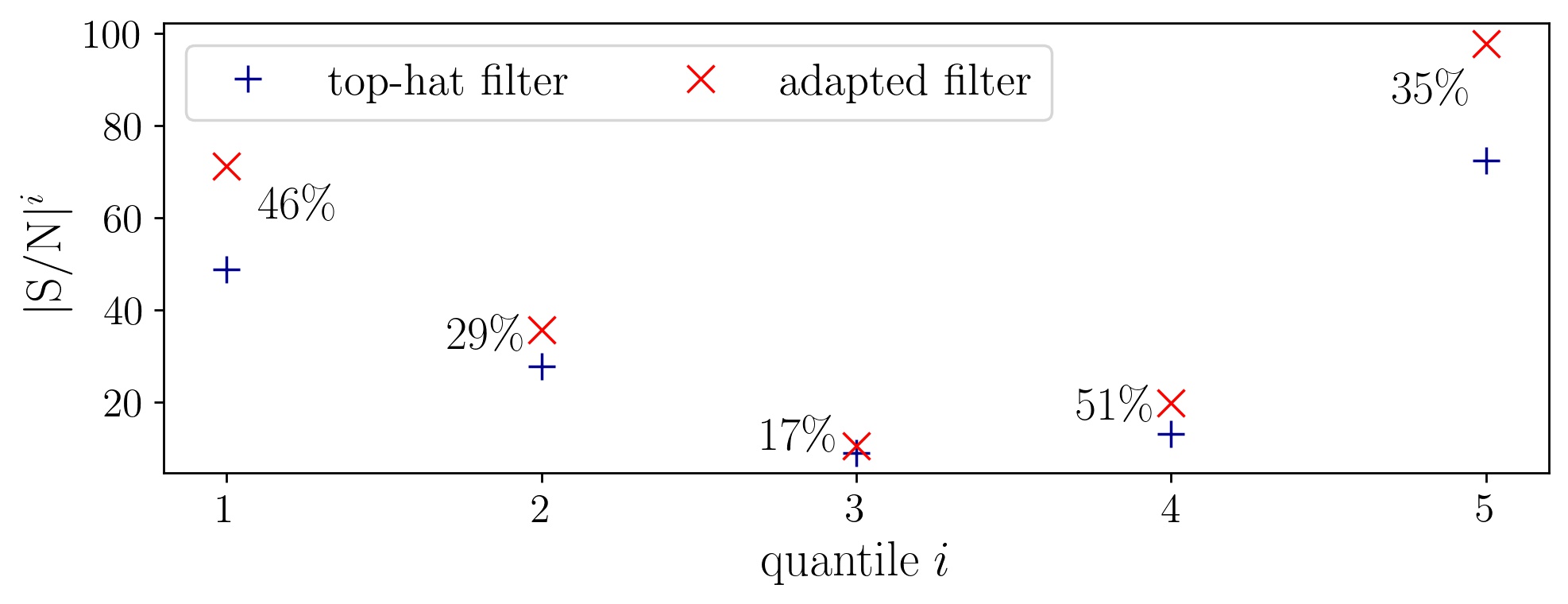}
    \caption{Comparison of the $|\nt{S}/\nt{N}|^i$ between the       adapted filter and a top-hat filter of size $\theta_\nt{th}=6'$ calculated with Eq.~\eqref{S/N_i} for the quantiles $i$. The uncertainties of the S/N are calculated with a jackknife estimator resampling the S/N data 64 times, removing at each draw one of the $\gamma_\nt{t}$ measurement and are below $2\%$. The values next to the points in the plot are the relative differences.}
    \label{SLICS_S2N}
\end{figure}

\section{$N_{\nt{ap}}$ versus $M_{\nt{ap}}$}
\label{MapNap_sec}

After deriving the adapted filter and specifying the top-hat filter size, we want to test our expectation that the adapted filter yields a better correlation between the galaxy and total matter distribution. For this analysis, we make use of 25 light cones from SLICS with a non-linear bias model and 25 light cones from the fiducial cosmology of cosmo-SLICS with a linear bias model, where we expect that for the latter the correlation is stronger since $n(\boldsymbol{\theta})\propto \kappa(\boldsymbol{\theta})$ here. For both models we calculate the aperture number with Eq.~\eqref{Nap} and the aperture mass with Eq.~\eqref{MapQ} for all pixels with the corresponding adapted filters and top-hat filters, where $\theta_\nt{th}=6'$ and $\theta_\nt{max}=30'$. For the aperture number we sum, as before, the foreground (lens) galaxies in squares of size 1\,arcmin$^2$, and for the aperture mass we average the ellipticities of background (source) galaxies in squares of size 1\,arcmin$^2$. Although we would expect similar relative correlation coefficients if we included shape noise in the shear estimates, we opted for the noise-free estimate to be closer to the true correlation coefficient. The results for both filter pairs are shown in Fig.~\ref{MapNap}, where the upper panels corresponds to the non-linear bias model (SLICS) and the lower panels to the linear bias model (fiducial cosmology from cosmo-SLICS). The correlation coefficient specified in the upper left corner of each panel is determined as
\begin{equation}
\rho = \frac{\left\langle \left(M_{\nt{ap}}(\boldsymbol{\theta})-\langle M_{\nt{ap}}\rangle\right)\left(N_{\nt{ap}}(\boldsymbol{\theta})-\langle N_{\nt{ap}}\rangle\right)\right\rangle }{\sqrt{\left\langle \left(M_{\nt{ap}}(\boldsymbol{\theta})-\langle M_{\nt{ap}}\rangle\right)^2\right\rangle \left\langle\left(N_{\nt{ap}}(\boldsymbol{\theta})-\langle N_{\nt{ap}} \rangle\right)^2\right\rangle}} \quad ,
\label{rho}
\end{equation}
where $\langle...\rangle$ refers to the ensemble average over all pixel positions $\boldsymbol{\theta}$ of the 25 light cones\footnote{Due to the periodic boundary effects of the FFT we do not consider the outer 30' margins.}. The higher the correlation factor $\rho$ is, the better the galaxy number density field traces the underlying matter field. As expected, the adapted filter yields a better correlation as seen in the correlation coefficient $\rho$, which is $\sim20\%$ higher for the adapted filter. Furthermore, it is seen that for the linear-bias model $\rho$ is $\sim 10\%$ higher. 

\begin{figure}[!htbp]
\centering
\begin{minipage}{\linewidth}
\includegraphics[width=\linewidth]{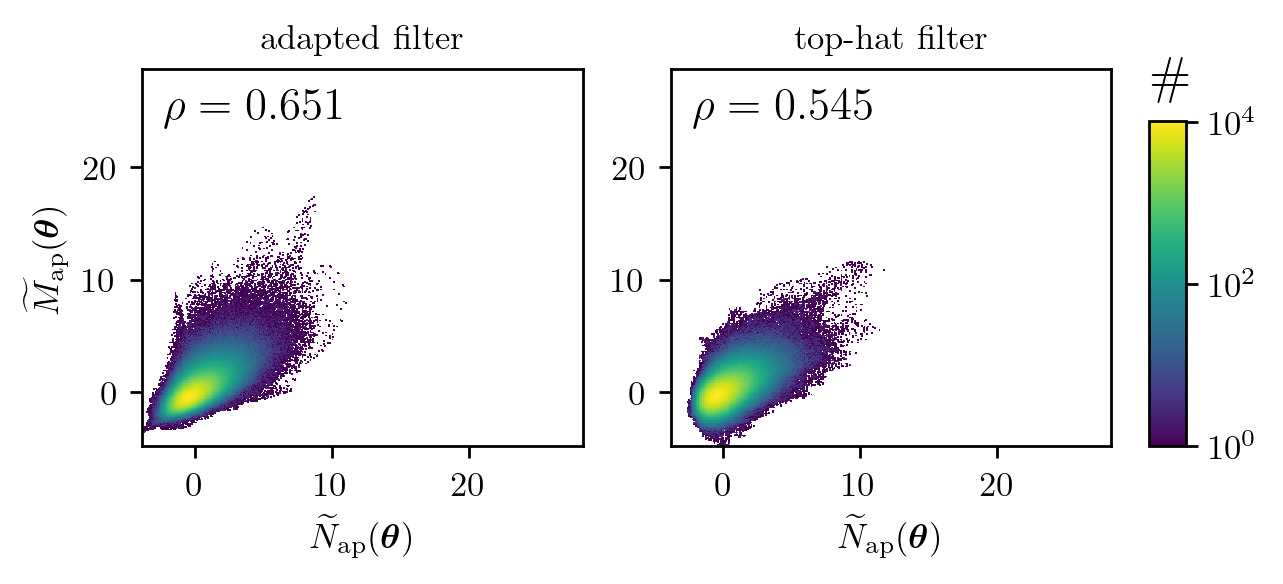}
\end{minipage}
\begin{minipage}{\linewidth}
\includegraphics[width=\linewidth]{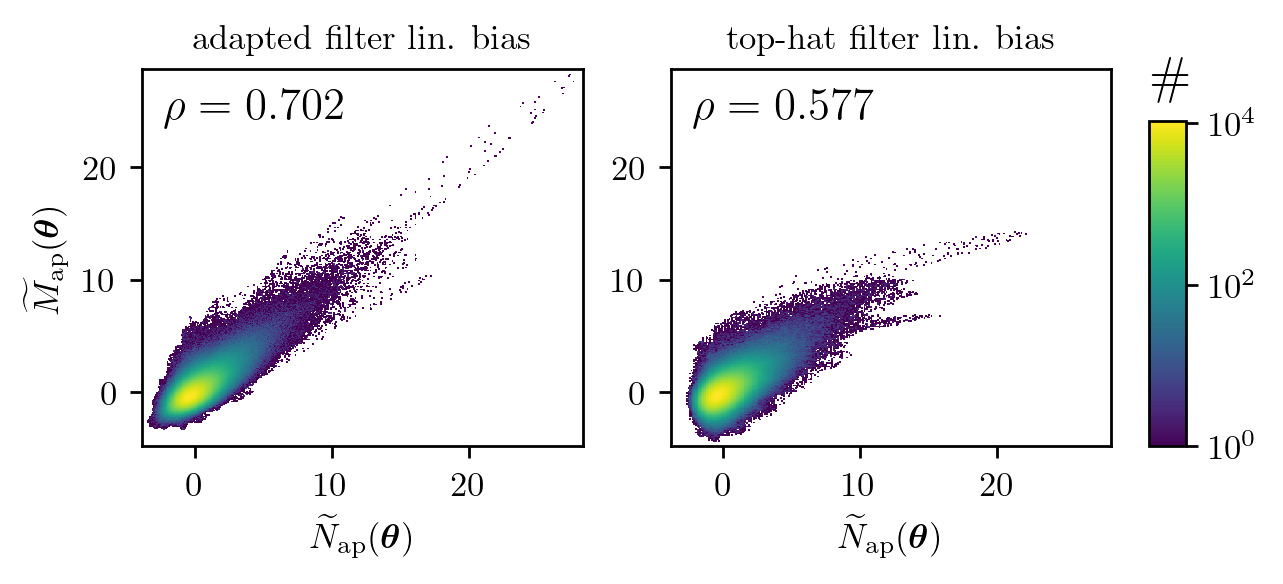}
\end{minipage}
\caption{Pixel-by-pixel $M_{\nt{ap}}(\boldsymbol{\theta})$ vs. $N_{\nt{ap}}(\boldsymbol{\theta})$ comparison for a non-linear bias model (upper panels) and linear bias model (lower panels). The aperture mass and number are calculated except for the outer margins for each individual pixel, which is different to Sect.~\ref{determinationTophat} where $M_{\nt{ap}}^i$ is calculated from shear profiles of specific quantiles. To ease the comparison between $M_{ \nt{ap}}(\boldsymbol{\theta})$ and $N_{ \nt{ap}}(\boldsymbol{\theta})$ we re-scaled $ M_{ \nt{ap}}(\boldsymbol{\theta})\rightarrow{\widetilde{M}_{\nt{ap}}(\boldsymbol{\theta}) := (M_{ \nt{ap}}(\boldsymbol{\theta})-\langle M_{\nt{ap}}\rangle)/\sqrt{\langle (M_{ \nt{ap}}(\boldsymbol{\theta})-\langle M_{\nt{ap}} \rangle)^2\rangle}}$, correspondingly $N_{ \nt{ap}}(\boldsymbol{\theta}) \rightarrow \widetilde{N}_{ \nt{ap}}(\boldsymbol{\theta})$, where  $\langle...\rangle$ is the ensemble average over all pixel positions $\boldsymbol{\theta}$. This re-scaling does not affect the correlation coefficient $\rho$ shown in the upper left corner of each panel. The panels on the left-hand side correspond to the adapted filter, and those on right-hand side to the top-hat filter. For both bias models, the adapted filter yields a stronger correlation, computed with Eq.~\eqref{rho}.}
    \label{MapNap}
\end{figure}

\section{Sensitivity to constrain cosmological parameters}
\label{CosCon}

\begin{figure}[!htbp]
\centering
    \includegraphics[width=\linewidth]{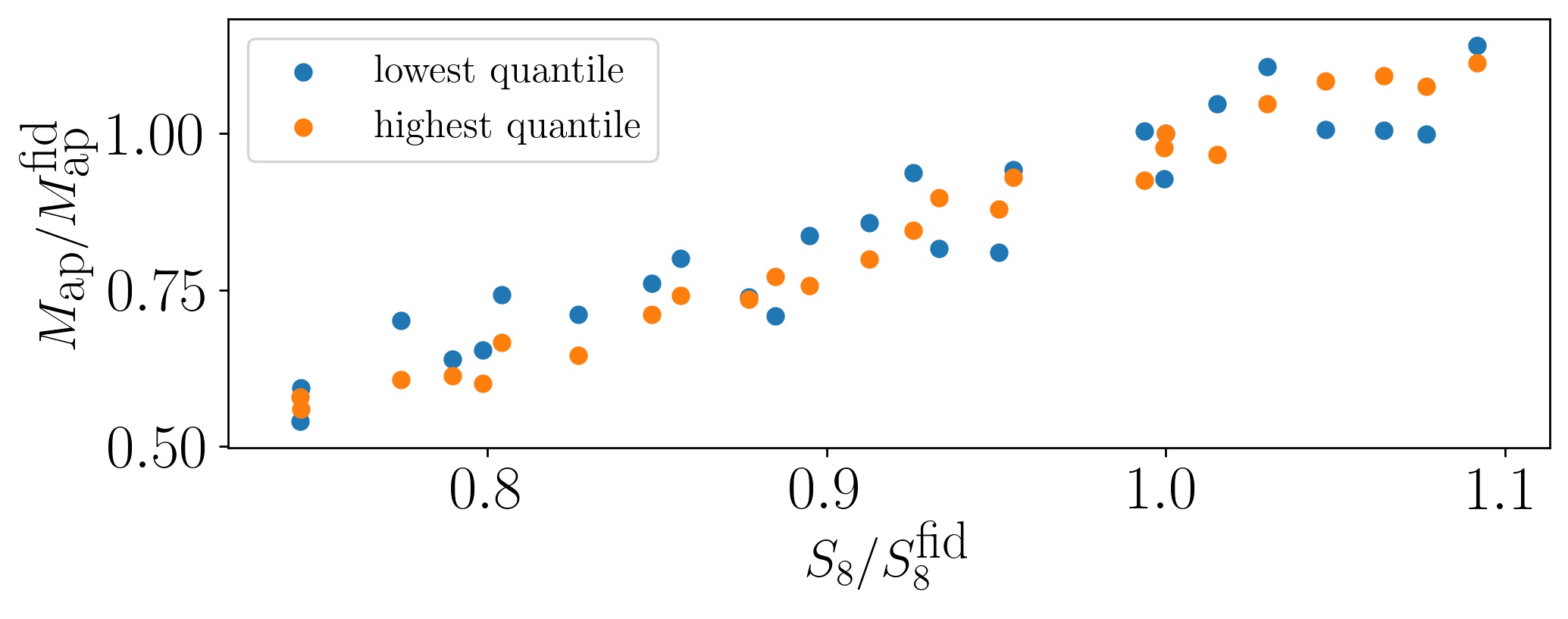}
    \caption{Comparison between $M_\nt{ap}$ and $S_8$ for the highest and lowest quantile from $N_\nt{ap}$ with the adapted filter of all cosmologies given in cosmo-SLICS. These two quantities are strongly correlated which indicates that $M_\nt{ap}$ is a useful cosmological probe. For the fiducial case of $S_8 = 0.8231$ the blue and the orange dot are both at $M_\nt{ap}/M_\nt{ap}^\nt{fid}=1$, so that you see only one orange dot since they are on top of each other.}
    \label{Map_S8}  
\end{figure}

\begin{figure*}[!htbp]
\sidecaption
    \includegraphics[width=12cm]{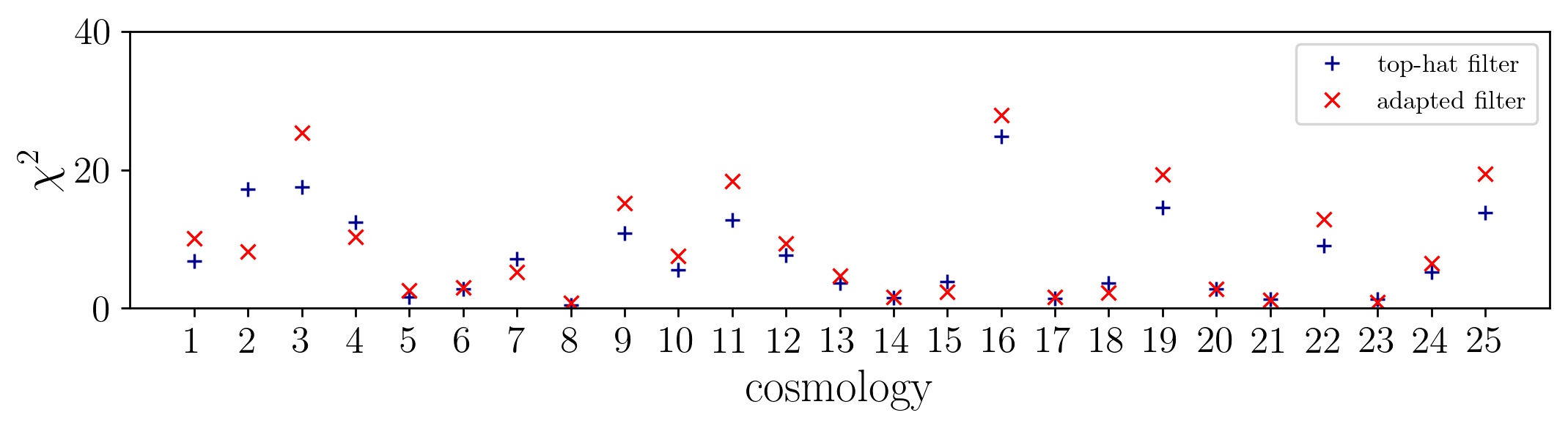}
    \caption{Comparison of the $\chi^2$ for all 25 cosmologies between
      the two filters, where the blue plus signs represent the top-hat filter
      and the red crosses correspond to the adapted filter. The
      parameters of the 25 cosmological model are shown in
      Table~\ref{cos_overview}.}
    \label{cos_filter_com}
\end{figure*}

\begin{figure*}[ht]
\centering
\begin{subfigure}{\columnwidth}
\includegraphics[width=\linewidth]{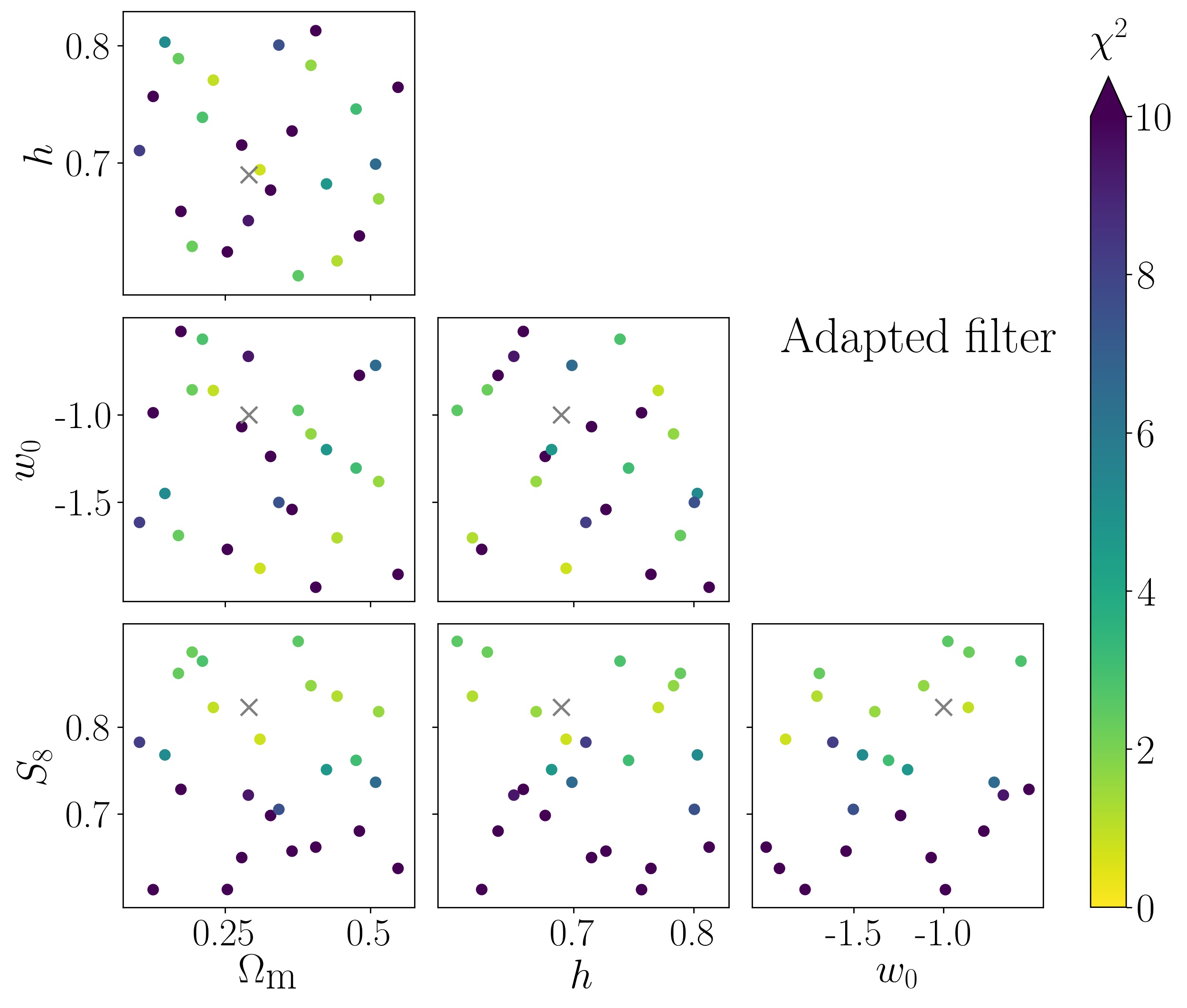}
\caption{Cosmo-SLICS: adapted filter}
\label{parametrespace:a} 
\end{subfigure}
\begin{subfigure}{\columnwidth}
\includegraphics[width=\linewidth]{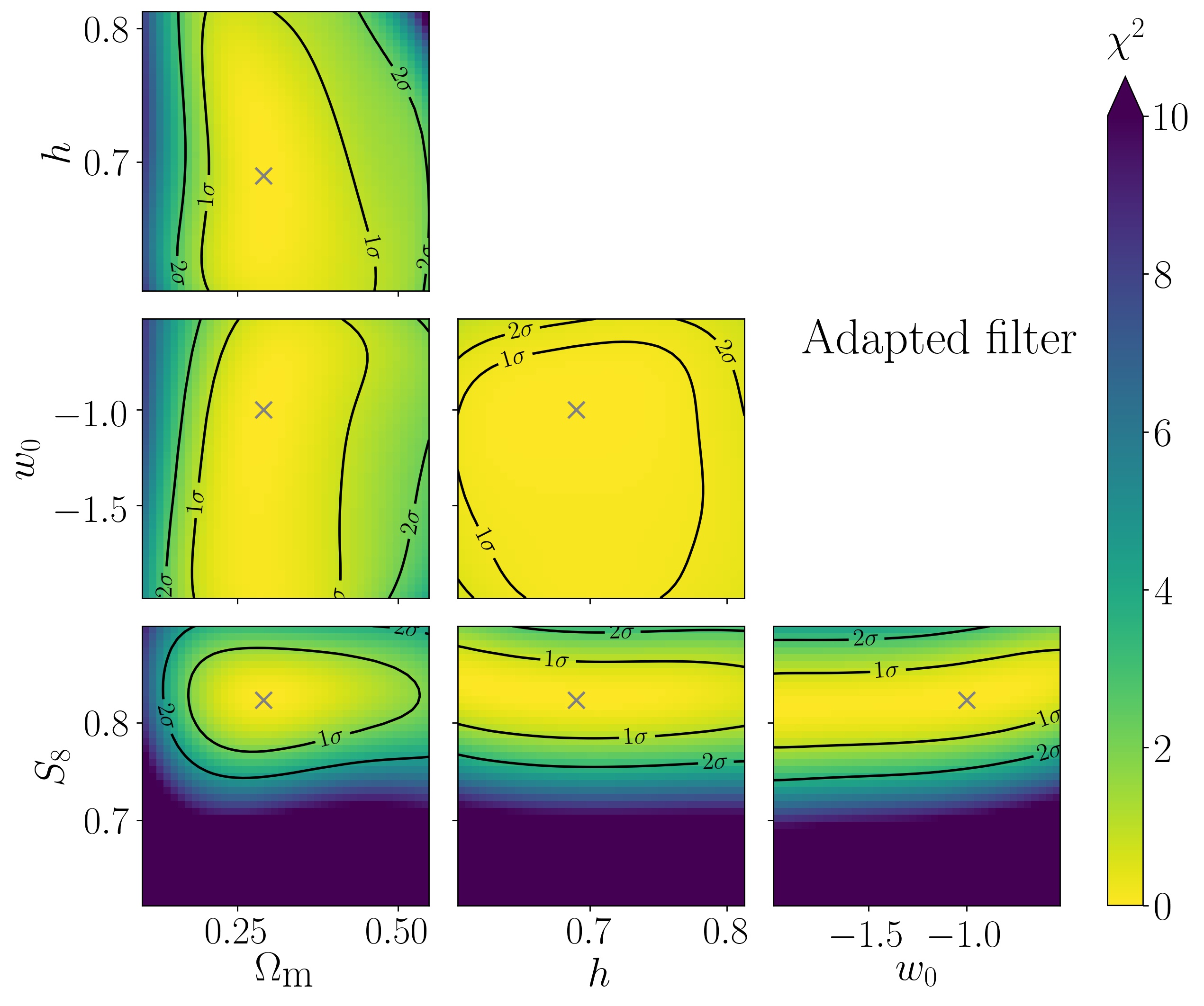}
\caption{GPRE: adapted filter}
\label{parametrespace:b}
\end{subfigure}
\begin{subfigure}{\columnwidth}
\includegraphics[width=\linewidth]{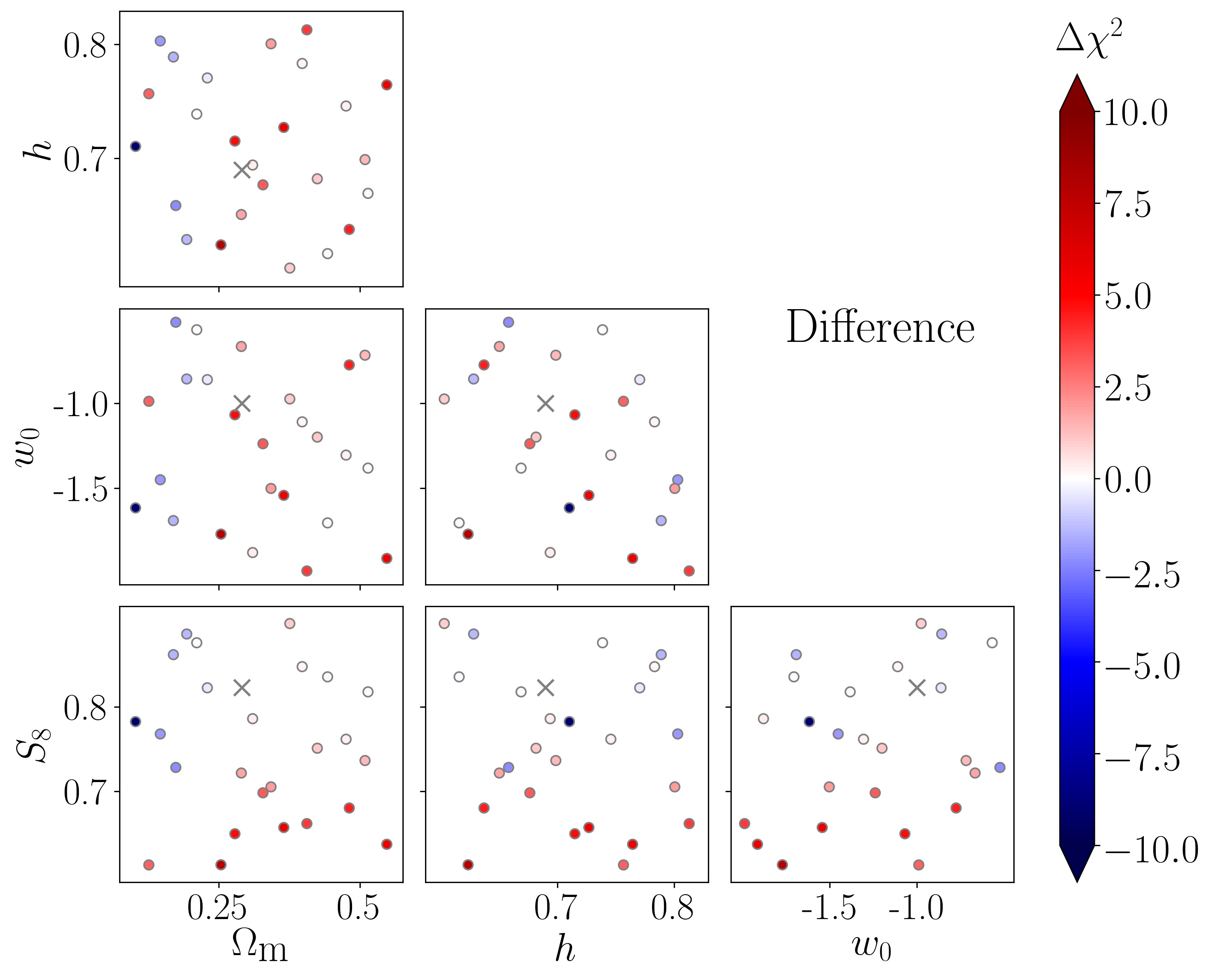}
\caption{Cosmo-SLICS: difference}
\label{parametrespace:c}
\end{subfigure}
\begin{subfigure}{\columnwidth}
\includegraphics[width=\linewidth]{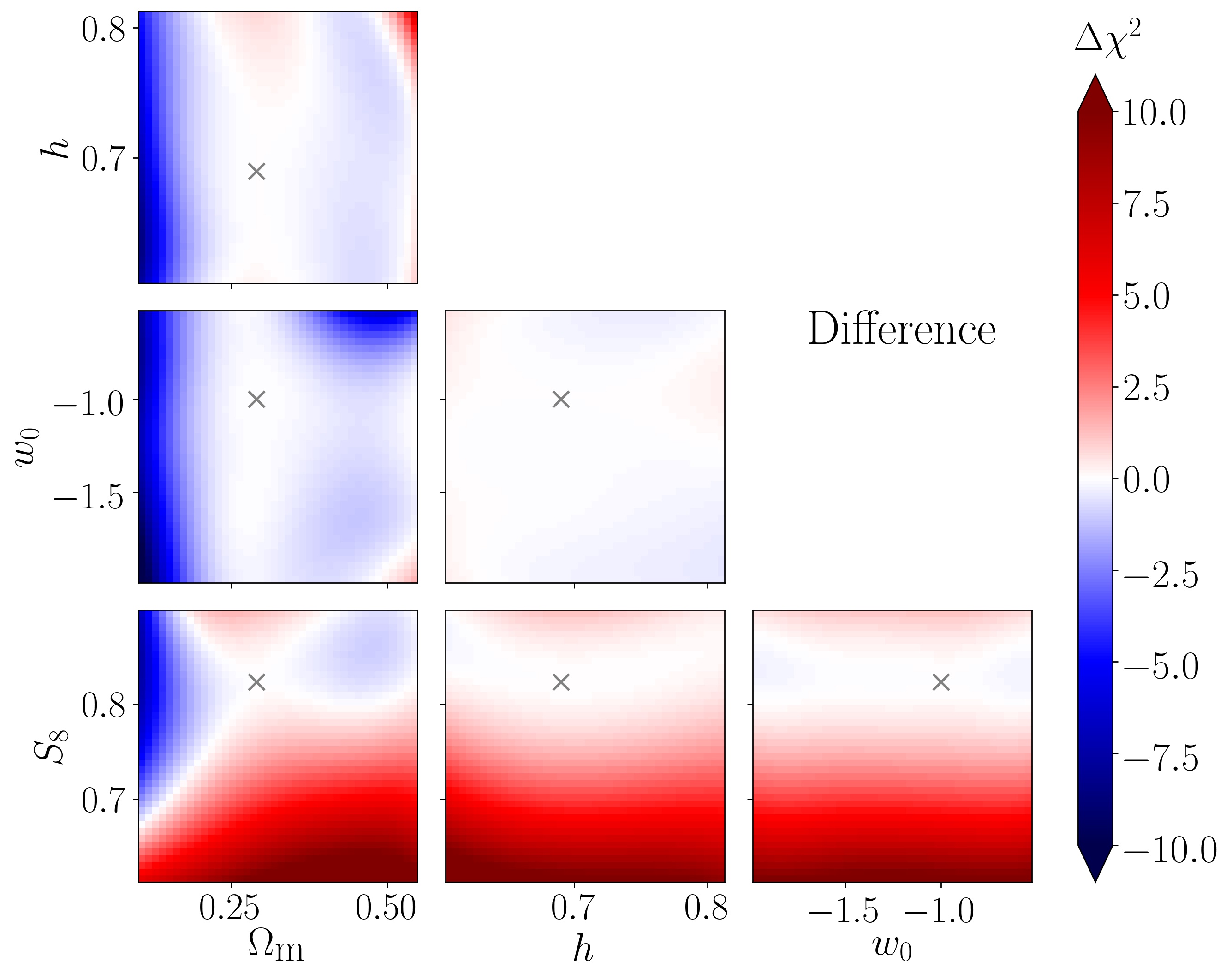}
\caption{GPRE: difference}
\label{parametrespace:d}
\end{subfigure}
\caption{Cosmological parameter space, where the colour in the upper
  panels indicates the $\chi^2$ corresponding to the adapted filter and in the lower panels to $\Delta \chi^2 = \chi^2_{\nt{ad}}-\chi^2_{\nt{th}}$. The $\chi^2$ of the left-hand side are determined with the cosmo-SLICS data and on the right-hand side with the flexible Gaussian process regression emulator. The grey cross marks the fiducial cosmology. One should not compare the right-hand side with the left-hand side directly, since in each node on the left, all four cosmological parameters are varied, whereas on the right, only two of the parameters are varied and the other two are fixed to the fiducial cosmology.}
\end{figure*}

\begin{figure}[!htbp]
\centering
\begin{minipage}{\linewidth}
\includegraphics[width=\linewidth]{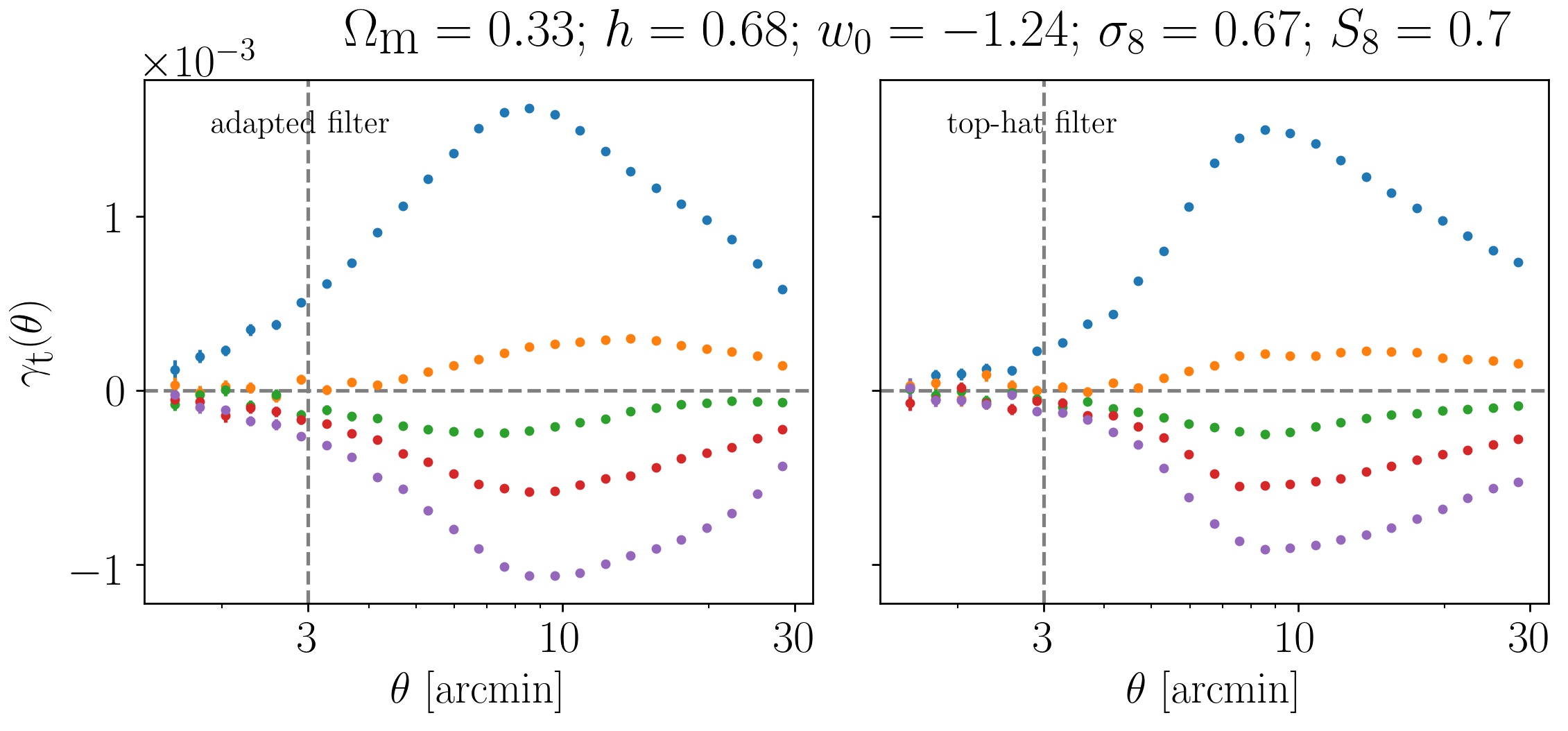}
\end{minipage}
\begin{minipage}{\linewidth}
\includegraphics[width=\linewidth]{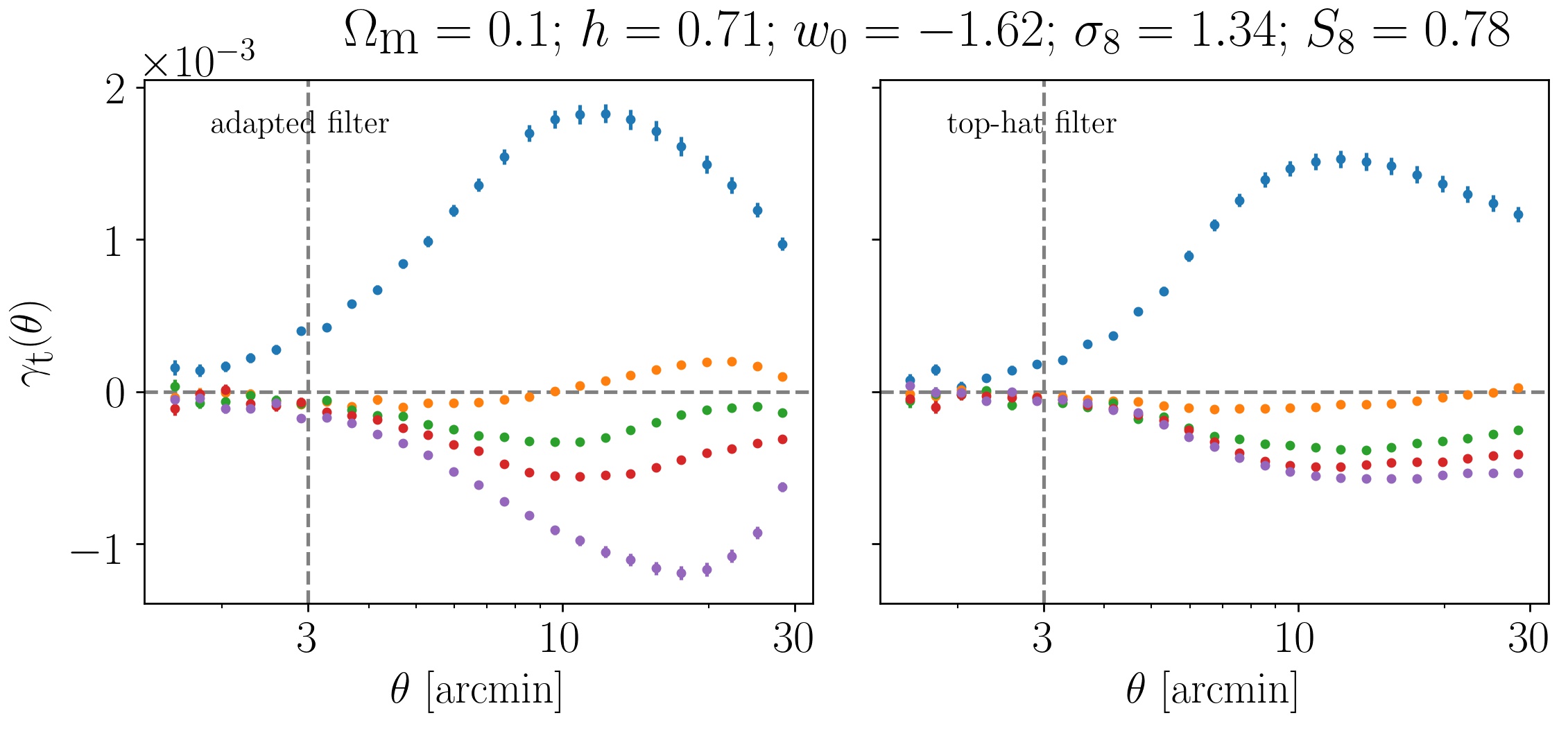}
\end{minipage}
\caption{Tangential shear profiles, $\gamma_\textrm{t}$, for two different cosmologies for the adapted filter on the left-hand side and the top-hat filter on the right-hand side. The uncertainties are the standard deviation on the mean determined with the 50 realisations.}
    \label{shear_cos}
\end{figure}

In this section, we investigate the sensitivity of the adapted and top-hat filters to varying cosmological parameters by use of the cosmo-SLICS, based on the aperture mass of Eq.~\eqref{MapQ_polar}. As seen in Fig.~\ref{Map_S8} for the highest and lowest quantile, $M_\nt{ap}$ and $S_8$ have a strong correlation, which indicates that $M_\nt{ap}$ is suitable as a metric for the comparison of different cosmologies.

For each of the 50 realisations per cosmology we first compute the aperture number with the two different filters and the \textsc{treecorr} $\gamma_{\rm t}$ profiles with shape noise of the five quantiles\footnote{Each quantile corresponds to one of the five sub-areas of the aperture number as in Sect.~\ref{determinationTophat}.}. Afterwards, we calculated an aperture mass $M_\nt{ap}^i$ by use of Eq.~\eqref{MapQ_polar} with the shear profiles of each realisation $n$. With these aperture masses we determine for each quantile in each cosmology an averaged aperture mass $\langle M_\nt{ap}^i \rangle$, where we average over the 50 realisation of each cosmology. Additionally we calculate one $5\times5$ covariance matrix for each filter (adapted and top-hat) from the shear profiles of the 50 fields for the  fiducial cosmology, which captures the correlation between the individual quantiles as
\begin{equation}
 C_{\nt{fid}}^{ij} = \frac{1}{50-1} \sum_{n=1}^{50} (M_{\nt{ap},n}^i-\langle M_\nt{ap}^i \rangle )(M_{\nt{ap},n}^j-\langle M_\nt{ap}^j \rangle) \quad ,
 \label{cov_cs}
 \end{equation}
 where $i$ and $j$ indicate the individual quantile and subscript 
'fid' indicates the fiducial cosmology. Using these quantities, we calculate for each cosmology with cosmological parameters $\mathbf{x}$ a $\chi^2$ as a measure of the deviation from the fiducial cosmology as
\begin{equation}
\chi^2(\mathbf{x}|\mathbf{M}_{\nt{ap}}^\nt{fid},C_{\nt{fid}}) =  \left[\mathbf{M}_{\nt{ap}}^\nt{fid}-\mathbf{M}_\nt{ap}(\mathbf{x})\right]^{\top} C_{\nt{fid}}^{-1}\left[\mathbf{M}_{\nt{ap}}^\nt{fid}-\mathbf{M}_\nt{ap}(\mathbf{x})\right]  ,
\label{chisquare}
\end{equation}
where $\mathbf{M}_\nt{ap}$ is the vector of the averaged amplitudes $\langle M_\nt{ap}^i \rangle$ of the five quantiles of the respective cosmology, with large $\chi^2$ values corresponding to deviations that are easier to detect.

The resulting $\chi^2$-values for the 25 cosmologies are displayed in
Fig.~\ref{cos_filter_com}, in which the $\chi^2$ for both filters are
compared to each other. It can be seen that the top-hat filter performs slightly better for some cosmologies with a low $\chi^2$. But for most cases, the adapted filter performs better in distinguishing between different cosmologies.

In order to investigate the sensitivity of the DSS to cosmological parameters in more detail, we display the two-dimensional parameter space in Fig.~\ref{parametrespace:a}, where the colour represents the $\chi^2$ of the analysis with the adapted filter. We see that the DSS is particularly powerful to distinguish between different values of $S_8=\sigma_8 \sqrt{\Omega_{\nt{m}}/0.3}$. Unfortunately, the cosmo-SLICS set does not cover values of $S_8>0.9$, but we expect that the $\chi^2$ would further increase. In contrast, there is hardly any correlation between $\chi^2$ and $w_0$, so that this parameter cannot be well constrained by DSS without a tomographic analysis.

Returning to the comparison between the adapted and the top-hat filter, we show in Fig.~\ref{parametrespace:c} the two-dimensional parameter space, but encoding in colour
$\Delta \chi^2 = \chi^2_{\nt{ad}}-\chi^2_{\nt{th}}$. For most
cosmologies, the adapted filter performs better, or no significant
difference is seen, which is consistent with
Fig.~\ref{cos_filter_com}. Whereas for most parameter pairs no clear trend with $\Delta\chi^2$ is seen, a clear correlation is present for $S_8$: For small $S_8$ values, the adapted filter performs better, but for large $S_8$ and small $\Omega_{\nt{m}}$ values (i.e. large $\sigma_8$), the top-hat filter is more sensitive. High $\sigma_8$ values imply strong clustering of the matter distribution. As a consequence, the analysis with the top-hat filter has difficulties to correctly assign regions of the sky within the lowest four quantiles, resulting in shear profiles with quite similar amplitudes, as seen in the lower right panel of Fig.~\ref{shear_cos}. The adapted filter is less affected by this effect, and therefore, $\chi^2$, which is a measure of the deviation to the fiducial cosmology, is larger for the top-hat filter than for the adapted filter (see Fig.~\ref{sigma8} for a visualisation of the $\Delta\chi^2$ in a $\sigma_8$-$\Omega_\nt{m}$ parameter space). Nevertheless, for all other cosmologies, the adapted filter is the better choice to distinguish different cosmologies.

\subsection*{Gaussian process regression emulator (GPRE)}

As all four cosmological parameters vary between the different cosmo-SLICS models, a comparison between the different cosmologies is non-trivial. To investigate the performance of the DSS with the two different filters on a continuous two-dimensional projected parameter space, we make use of a flexible GPRE described in Sect. A1 in \citet{Harnois-Deraps:2019} to emulate averaged tangential shears $\gamma_\nt{t}$ for various cosmologies. The training of the emulator for each individual quantile and for both filters is carried out with the 26 cosmo-SLICS cosmologies. In order to test the accuracy of the GPRE we apply the `leave-one-out' cross-validation method and show the results in Fig.~\ref{ACC_emu}. The shear profiles of the highest and lowest two quantiles can be predicted with an accuracy of generally better than 10$\%$. The shear profiles of the fourth and third quantile have a relative accuracy far worse than that, but this is not surprising since these quantiles have a very low shear signal. However, we checked that our results are robust with respect to including or excluding these two quantiles.

In order to produce smooth two-dimensional constraints on the four cosmological parameters, we vary two of the four parameters in 41 steps in the same range as the parameters were given in cosmo-SLICS and fixed the other two remaining parameters to the fiducial cosmology. Next, we calculate for each grid point the aperture masses $M_\nt{ap}^i$ from Eq.~\eqref{MapQ_polar} and $\chi^2$ from Eq.~\eqref{chisquare} as measures of the deviation of the predicted shear profiles from the predicted fiducial shear profiles. We emulate directly the averaged shear profiles, so that $\mathbf{M}_\nt{ap}$ in Eq.~\eqref{chisquare} is the vector of $M_\nt{ap}^i$ calculated with the emulated shear profiles. The covariance matrix employed is the one calculated with the 50 realisations from the fiducial cosmology of cosmo-SLICS by use of Eq.~\eqref{cov_cs}. The results for the individual parameter pairs are displayed  in Fig.~\ref{parametrespace:b}. As expected, the further we deviate from the fiducial cosmology the higher is the $\chi^2$. By inspecting the individual panels, we see that the $S_8$ and $\Omega_{\textrm{m}}$ parameters are well constrained. Furthermore, it is clearly visible that these two parameters are dominating the change in the shear profiles for all parameter pairs. This can be seen especially in the case when $S_8$ and $\Omega_{\textrm{m}}$ are fixed and $h$ or $w_0$ are varied, where the $\chi^2$ has a very weak gradient.

We also investigated the difference between the adapted and top-hat filters, seen in Fig.~\ref{parametrespace:d}. Around the fiducial cosmology, the $\chi^2$ values of both filters are indistinguishable, but as the trend of the 25 cosmo-SLICS nodes (Fig.~\ref{parametrespace:c}) already suggests, the analysis with top-hat is more sensitive for large $\sigma_8$ values, whereas the adapted filter is better for the remaining parameter regions.

Summarising this section, we find that the top-hat and adapted
filters perform similarly around the fiducial cosmology to differentiate cosmologies, but moving away from the fiducial cosmology the adapted filter is more constraining than the top-hat filter, with the exception of large $\sigma_8$ values.

\section{Summary and conclusion}
\label{Conclusion}

In this work, we constructed a pair of adapted filter functions for
the DSS, using ray tracing and a semi-analytic model galaxy population in the MS in an iterative process. Our new pair of filters is matched with respect to the aperture mass and galaxy number statistics. In other words, the adapted pair of filters measures the lensing convergence and the galaxy number density with the same angular weighting. Based on numerical weak lensing simulations, we confirmed our expectation that the correlation between galaxy number density and shear signal is higher with our adapted filter than for the top-hat filter. We verified that this result holds both for a linear and a non-linear galaxy bias model, using mock GAMA$\times$KV450 data constructed from the SLICS and the cosmo-SLICS weak lensing simulations.

Furthermore, we showed that the adapted filter is indeed a useful improvement for the DSS, by comparing it with the previously used top-hat filter of appropriate scale, using their resulting S/N in different sub-areas of the sky and their sensitivity to discriminate between different sets of cosmological parameters as metrics. These sub-areas are called quantiles of the aperture number field. For the S/N comparison, we made use of the $w$CDM SLICS simulation and showed that the adapted filter has a higher S/N for most quantiles. For comparing the sensitivity of both filters to different cosmologies, we employed the cosmo-SLICS, which is a suite of 26 different cosmologies with 50 realisations each. From the 50 realisations in each cosmology, we calculated a $\chi^2$ as a measure for the deviation from the fiducial cosmology. It turned out that both filters behave similarly near the fiducial cosmology, but that the adapted filter is more constraining in most regions of parameter space probed by cosmo-SLICS, except for very high values of $\sigma_8$ where the top-hat filter yielded higher deviation from the fiducial cosmology. In order to investigate the performance of the DSS with the two different filters on a continuous two-dimensional projected parameter space, we also made use of a flexible GPRE, which is a promising tool for future cosmological analyses. Both the S/N and the cosmological analyses lead to the conclusion that the adapted filter yields tighter cosmological constraints than the top-hat filter and should be employed in future DSS analyses.

As an outlook, it would be interesting to investigate the
arbitrariness of dividing the aperture number field into five quantiles. For instance, one could optimally combine the shear profiles or find a way to not bin the sky at all, as binning decreases the information content. Furthermore, the filter size used here has not been optimised and should also be varied. Our first attempt here to look into the usefulness of the new DSS to constrain cosmological parameters relied fully on numerical simulations, we aim to modify the analytical model derived by \citet{Friedrich:Gruen:2018} to account for the adapted filter, allowing for an alternative modelling option in an up-coming cosmological study, similar to the approach of \citet{Gruen:Friedrich:2018}.

\begin{acknowledgement}
We thank the anonymous referee for very constructive and fruitful comments. Furhter, we would like to thank Oliver Friedrich, Patrick Simon, Jan Luca von den Busch and Sven Heydenreich for useful discussions, and  Mike Jarvis for maintaining \textsc{treecorr}. The results in this paper are based on observations made with ESO Telescopes at the La Silla Paranal Observatory under programme IDs 177.A-3016, 177.A-3017, 177.A-3018 and 179.A-2004, and on data products produced by the KiDS consortium. The KiDS production team acknowledges support from: Deutsche  Forschungsgemeinschaft,  ERC,  NOVA  and  NWO-M  grants;  Target; the University of Padova, and the University Federico II (Naples). We acknowledge support from the European Research Council under grant numbers 770935 (HH) and 647112 (CH, JHD). PB acknowledges supported by the Deutsche Forschungsgemeinschaft, project SCHN342-13. H. Hildebrandt is supported by a Heisenberg grant of the Deutsche Forschungsgemeinschaft (Hi 1495/5-1). JHD is  supported  by  a STFC Ernest Rutherford Fellowship (project reference ST/S004858/1). SU is partly supported by the German Deutsche Forschungsgemeinschaft, DFG project numbers SL 172/1-1 and SCHN 342/13-1. VD acknowledges the Higgs Centre Nimmo Scholarship and the Edinburgh Global Research Scholarship. CH acknowledges support from the Max Planck Society and the Alexander von Humboldt Foundation in the framework of the Max Planck-Humboldt Research Award endowed  by  the  Federal  Ministry  of  Education  and  Research. Author contributions: all authors contributed to the development and writing of this paper. The authorship list is given in two groups: the lead authors (PB, PS), followed by an alphabetical group, which provided data central to this work, or contributed to the analysis.
\end{acknowledgement}

\bibliographystyle{aa}
\bibliography{cite}

\begin{appendix}
\section{Additional material}

\begin{table}[h!]
\centering
\caption{Overview of all the different cosmological parameters for the 26 cosmo-SLICS models, which are used in Sect.~\ref{CosCon} for the cosmological analysis.}
\begin{tabular}{c|ccccc}
 & $\Omega_\nt{m}$ & $h$ & $w_0$ & $\sigma_8$ & S$_8$ \\
\hline
fid & 0.2905 & 0.6898 & $-1.0000$ & 0.8364 & 0.8231 \\
1 & 0.3282 & 0.6766 & $-1.2376$ & 0.6677 & 0.6984 \\
2 & 0.1019 & 0.7104 & $-1.6154$ & 1.3428 & 0.7826 \\
3 & 0.2536 & 0.6238 & $-1.7698$ & 0.6670 & 0.6133 \\
4 & 0.1734 & 0.6584 & $-0.5223$ & 0.9581 & 0.7284 \\
5 & 0.3759 & 0.6034 & $-0.9741$ & 0.8028 & 0.8986 \\
6 & 0.4758 & 0.7459 & $-1.3046$ & 0.6049 & 0.7618 \\
7 & 0.1458 & 0.8031 & $-1.4498$ & 1.1017 & 0.7680 \\
8 & 0.3099 & 0.6940 & $-1.8784$ & 0.7734 & 0.7861 \\
9 & 0.4815 & 0.6374 & $-0.7737$ & 0.5371 & 0.6804 \\
10 & 0.3425 & 0.8006 & $-1.5010$ & 0.6602 & 0.7054 \\
11 & 0.5482 & 0.7645 & $-1.9127$ & 0.4716 & 0.6375 \\
12 & 0.2898 & 0.6505 & $-0.6649$ & 0.7344 & 0.7218 \\
13 & 0.4247 & 0.6819 & $-1.1986$ & 0.6313 & 0.7511 \\
14 & 0.3979 & 0.7833 & $-1.1088$ & 0.7360 & 0.8476 \\
15 & 0.1691 & 0.7890 & $-1.6903$ & 1.1479 & 0.8618 \\
16 & 0.1255 & 0.7567 & $-0.9878$ & 0.9479 & 0.6131 \\
17 & 0.5148 & 0.6691 & $-1.3812$ & 0.6243 & 0.8178 \\
18 & 0.1928 & 0.6285 & $-0.8564$ & 1.1055 & 0.8862 \\
19 & 0.2784 & 0.7151 & $-1.0673$ & 0.6747 & 0.6500 \\
20 & 0.2106 & 0.7388 & $-0.5667$ & 1.0454 & 0.8759 \\
21 & 0.4430 & 0.6161 & $-1.7037$ & 0.6876 & 0.8356 \\
22 & 0.4062 & 0.8129 & $-1.9866$ & 0.5689 & 0.6620 \\
23 & 0.2294 & 0.7706 & $-0.8602$ & 0.9407 & 0.8226 \\
24 & 0.5095 & 0.6988 & $-0.7164$ & 0.5652 & 0.7366 \\
25 & 0.3652 & 0.7271 & $-1.5414$ & 0.5958 & 0.6574 \\
\end{tabular}
\label{cos_overview}
\end{table}

\begin{figure}[!htbp]
\centering
    \includegraphics[width=\linewidth]{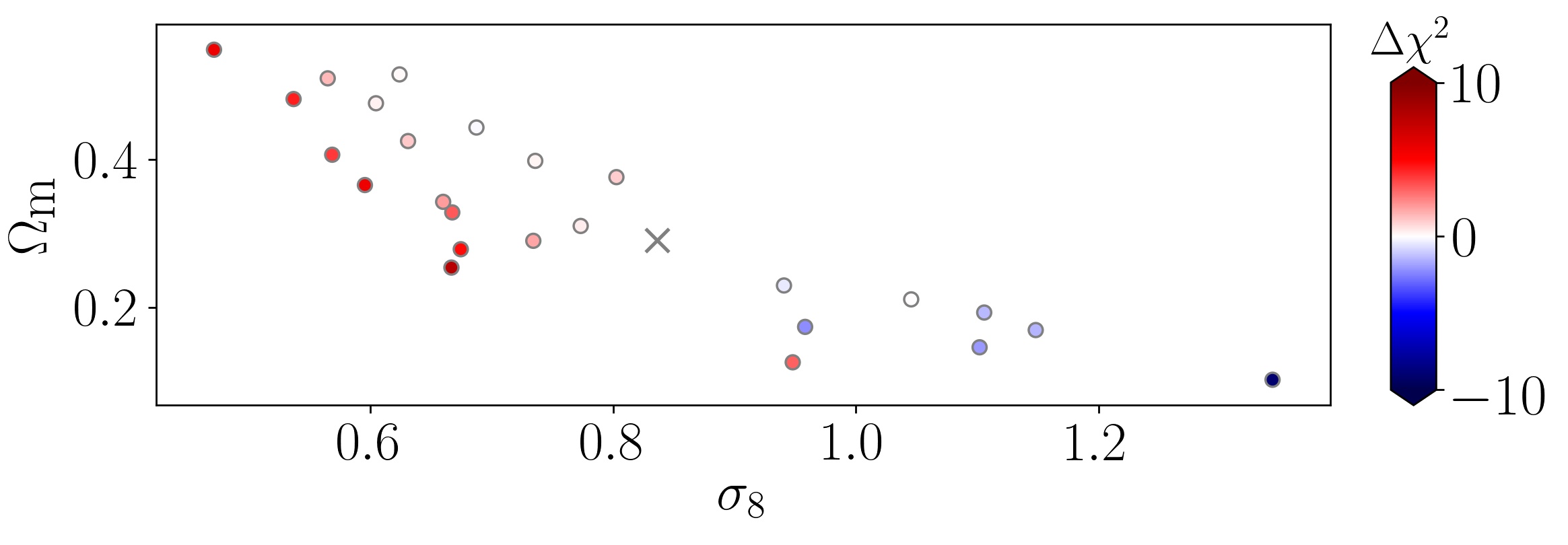}
    \caption{Cosmological parameter space $\sigma_8$-$\Omega_\nt{m}$, where the colour indicates $\Delta \chi^2 = \chi^2_{\nt{ad}}-\chi^2_{\nt{th}}$ of the 25 nodes of cosmo-SLICS determined in Sect.~\ref{CosCon}. It is clearly seen that for large $\sigma_8$ the analysis with the top-hat filter yields higher $\chi^2$. The grey cross indicates the fiducial cosmology.}
    \label{sigma8}
\end{figure}

\begin{figure}[!htbp]
\centering
\begin{minipage}{\linewidth}
\includegraphics[width=\linewidth]{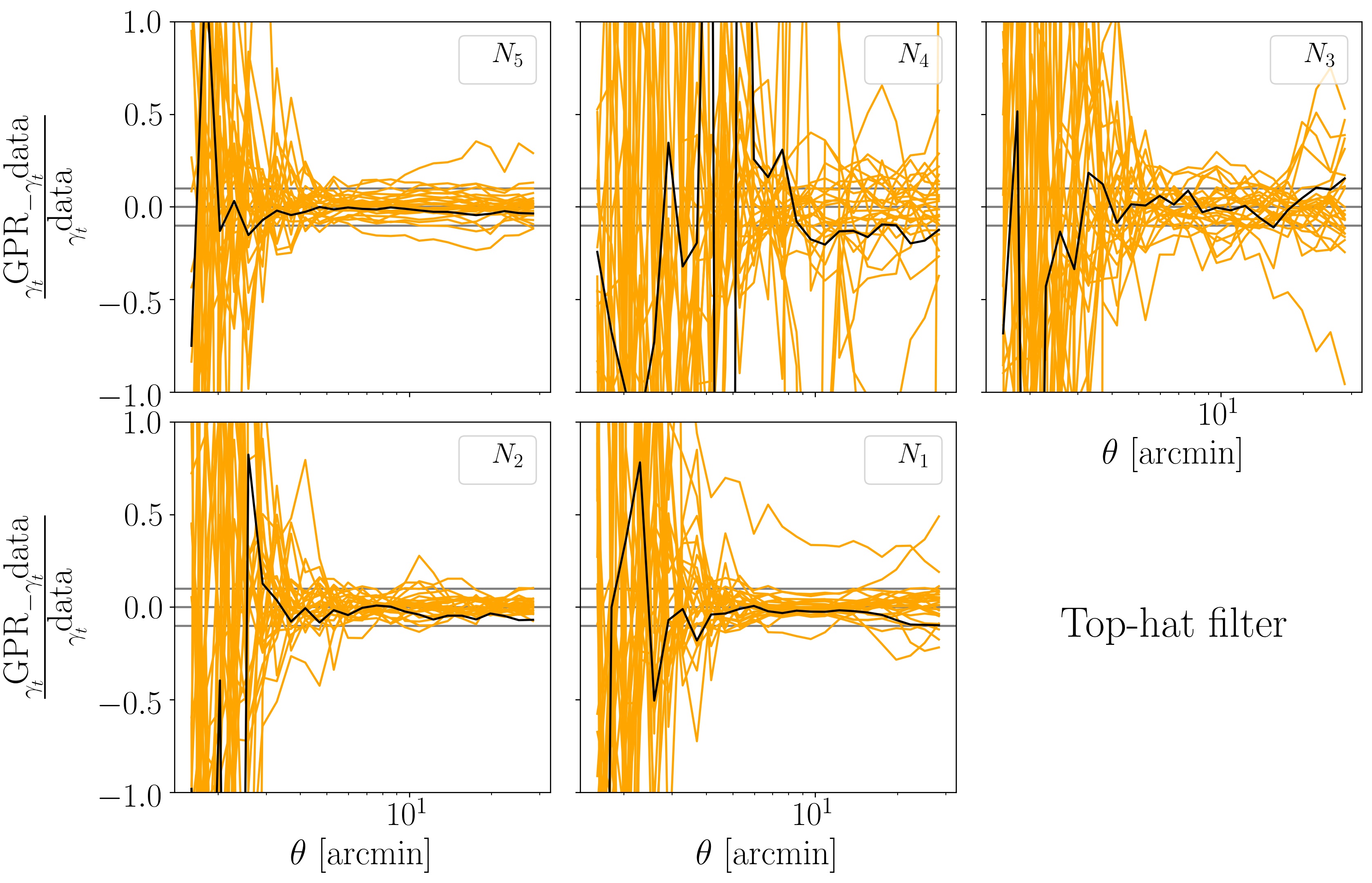}
\end{minipage}
\begin{minipage}{\linewidth}
\includegraphics[width=\linewidth]{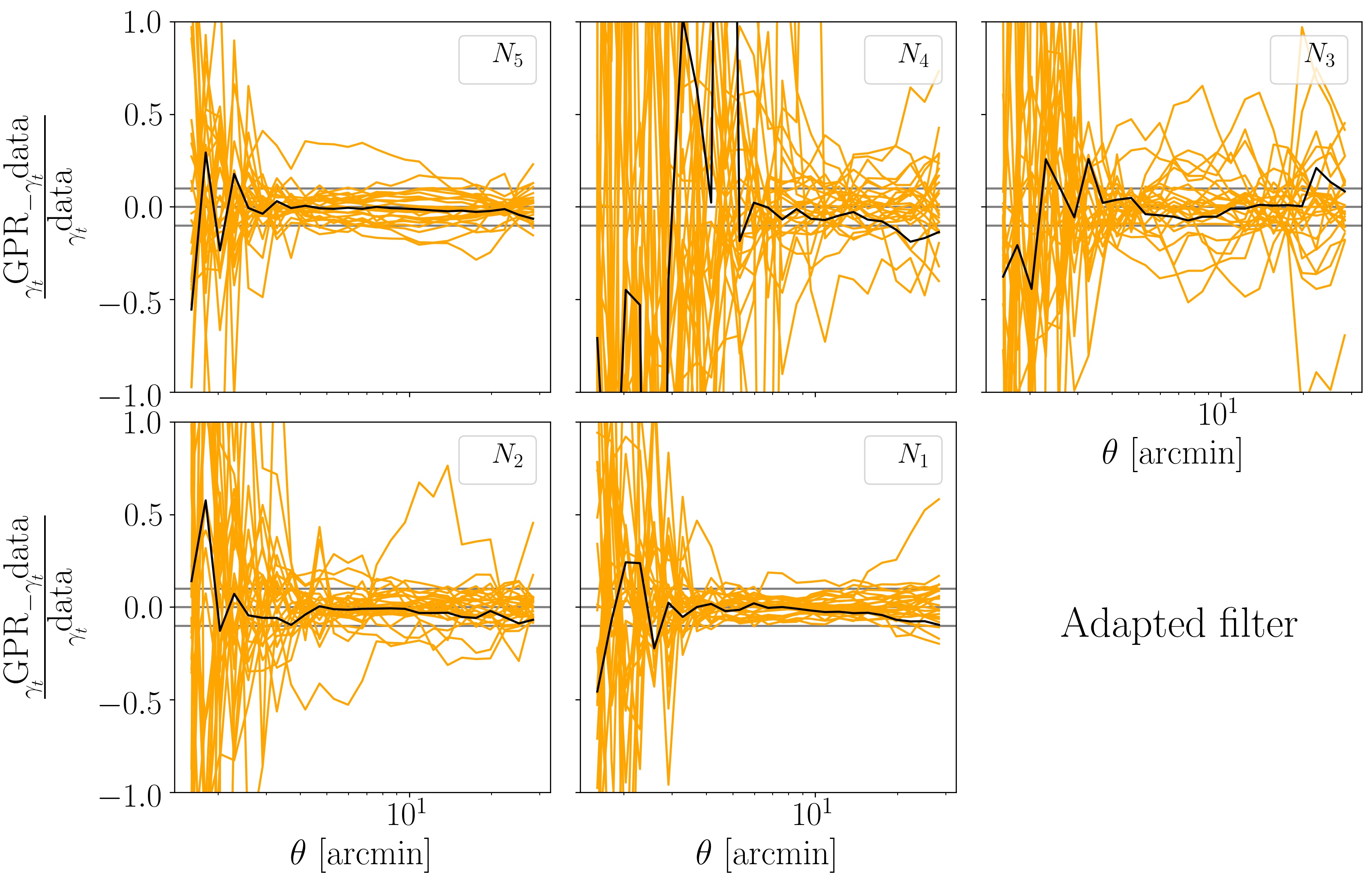}
\end{minipage}
      \caption{'Leave-one-out' cross-validation to test performance of accuracy of the GPRE, which is used in Sect.~\ref{CosCon} to investigate the performance of the DSS with the two different filters on a continuous two-dimensional projected parameter space. On the $y$-axis the relative difference between the predicted shear profile of one cosmology if the emulator is trained exclusively by the remaining cosmologies and the corresponding shear profile which the emulator tries to emulate. The black lines here are correspond to the fiducial case. The quantiles N$_4$ and N$_3$ are quite inaccurate, but the other quantiles are of the 10$\%$ level accurate, which are indicated with horizontal grey lines.}
    \label{ACC_emu}
\end{figure}

\end{appendix}

\end{document}